\documentclass[msom]{template/template}

\DoubleSpacedXI

\usepackage{caption}
\usepackage[utf8]{inputenc}
\DeclareUnicodeCharacter{1F3AF}{\target}
\DeclareRobustCommand\target{%
  \unskip\nobreak\thinspace\textemdash\allowbreak\thinspace\ignorespaces}

\usepackage[font={footnotesize, rm, bf}, labelsep=quad, labelfont = {footnotesize, rm}]{caption}
\usepackage{subcaption}
\usepackage{filecontents}
\usepackage[noend]{algpseudocode}
\usepackage{algorithm}

\usepackage{pifont}
\usepackage{hyperref}
\usepackage{tikz}
\usetikzlibrary{calc, shapes, arrows}
\usepackage{comment}
\usepackage{accents}
\usepackage{multirow}
\usepackage{tabularx}
\usepackage{booktabs}

\usepackage{pgfplots}
  \usepgfplotslibrary{statistics}
  \pgfplotsset{compat=1.8}
  \usepgfplotslibrary{fillbetween}
  \pgfmathdeclarefunction{fpumod}{2}{%
    \pgfmathfloatdivide{#1}{#2}%
    \pgfmathfloatint{\pgfmathresult}%
    \pgfmathfloatmultiply{\pgfmathresult}{#2}%
    \pgfmathfloatsubtract{#1}{\pgfmathresult}%
    \pgfmathfloatifapproxequalrel{\pgfmathresult}{#2}{\def\pgfmathresult{5}}{}%
  }

\definecolor{Tblue}{HTML}{0065BD}
\definecolor{TUMgray}{HTML}{CCCCCC}
\definecolor{TUMdarkgray}{HTML}{808080}
\definecolor{TUMblue2}{HTML}{005293}
\definecolor{TUMblue3}{HTML}{003359}
\definecolor{TUMalblack}{HTML}{333333}
\definecolor{TUMorange}{HTML}{E37222}

\newenvironment{customlegend}[1][]{%
  \begingroup
  \csname pgfplots@init@cleared@structures\endcsname
  \pgfplotsset{#1}%
}{%
  \csname pgfplots@createlegend\endcsname
  \endgroup
}%

\def\addlegendimage{\csname pgfplots@addlegendimage\endcsname}

\usepackage[bb=boondox]{mathalfa}
\usepackage{booktabs}
 \usepackage{wrapfig}
 \usepackage{enumitem}
 \usepackage[utf8]{inputenc}
\usepackage{amssymb}
\usepackage{amsmath}

\usepackage{dsfont}
\usepgfplotslibrary{statistics}
\usepackage{array}

\definecolor{TUMblue}{RGB}{0,57,166}

\newcolumntype{L}[1]{>{\raggedright\let\newline\\\arraybackslash\hspace{0pt}}m{#1}}
\newcolumntype{C}[1]{>{\centering\let\newline\\\arraybackslash\hspace{0pt}}m{#1}}
\newcolumntype{R}[1]{>{\raggedleft\let\newline\\\arraybackslash\hspace{0pt}}m{#1}}

\usepackage{natbib}
 \bibpunct[, ]{(}{)}{,}{a}{}{,}%
 \def\bibfont{\small}%
\TheoremsNumberedThrough   

\EquationsNumberedThrough  

\MANUSCRIPTNO{} 
\renewcommand{\qed}{$\square$}

\begin{document}


 \RUNAUTHOR{
Bauerhenne, Kolisch, and Schulz}

\RUNTITLE{Robust Appointment Scheduling with Waiting Time Guarantees}

\RUNTITLE{Robust Appointment Scheduling with Waiting Time Guarantees}

 \TITLE{Robust Appointment Scheduling with Waiting Time Guarantees}

\ARTICLEAUTHORS{%
\AUTHOR{
Carolin Bauerhenne}
\AFF{School of Management, Technical University of Munich, Munich, Germany, \EMAIL{carolin.bauerhenne@tum.de} }
\AUTHOR{Rainer Kolisch}
\AFF{School of Management, Technical University of Munich, Munich, Germany, \EMAIL{rainer.kolisch@tum.de}}
\AUTHOR{Andreas S. Schulz}
\AFF{School of Computation, Information, and Technology \& School of Management, Technical University of Munich, Munich, Germany, \EMAIL{andreas.s.schulz@tum.de}}
} 

\ABSTRACT{%
\textbf{Abstract.} \textbf{\textit{Problem definition:}} 
Appointment scheduling problems under uncertainty encounter a fundamental trade-off between cost minimization and customer waiting times. Most existing studies address this trade-off using a weighted sum approach, which puts little emphasis on individual waiting times and, thus, customer satisfaction. In contrast, we study how to minimize total cost while providing waiting time guarantees to all customers.
\textbf{\textit{Methodology/results:}} Given box uncertainty sets for service times and no-shows, we introduce the Robust Appointment Scheduling Problem with Waiting Time Guarantees. We show that the problem is $\mathcal{NP}$-hard in general and introduce a mixed-integer linear program that can be solved in reasonable computation time. For special cases, we prove that polynomial-time variants of the well-known Smallest-Variance-First sequencing rule and the Bailey-Welch scheduling rule are optimal. Furthermore, a case study with data from the radiology department of a large university hospital demonstrates that the approach not only guarantees acceptable waiting times but, compared to existing robust approaches, may simultaneously reduce costs incurred by idle time and overtime. \textbf{\textit{Managerial implications:}} This work suggests that limiting instead of minimizing customer waiting times is a win-win solution in the trade-off between customer satisfaction and cost minimization. Additionally, it provides an easy-to-implement and customizable appointment scheduling framework with waiting time guarantees.
}%


\maketitle
\section{Introduction}\label{intro} 
Excessive waiting times negatively affect customer satisfaction, perceived service quality, and loyalty in various services, such as call centers, public transportation, and health care \citep{huang1994patient, taylor1994assessment}.
A customer's tolerance for waiting, which separates acceptable from excessive waiting times, has been shown to range from 37 minutes for a majority of patients in an outpatient clinic to 11 seconds for most users to load a website \citep{bouch2000quality, moschis2003influences, hill2005impact}. The tolerance depends amongst others on the perceived and expected waiting time and can be influenced by many effective interventions, e.g., \cite{kumar1997impact, davis1998disconfirmation, bielen2007waiting}. Meanwhile, there is a lack of appointment scheduling approaches to control actual waiting times.

Appointment scheduling is a multi-criteria problem that typically involves uncertainty in service times and other attributes. In particular, appointment scheduling problems face a fundamental trade-off between waiting times and capacity utilization: Shorter waiting times come at the cost of longer idle times, and vice versa. In the literature, this trade-off has been modeled predominantly by minimizing a weighted sum of waiting times, idle times, and overtime \citep{ahmadi2017outpatient}. However, it has long been known that such weighted sum approaches can lead to unbalanced and excessive waiting times for some customers \citep{cayirli2003outpatient}, and in combination with individual no-show predictions even to racial disparity \citep{samorani2020overbooked}.

This paper proposes a new approach to appointment scheduling that considers waiting (in)tolerance. In particular, we study the following robust appointment scheduling and sequencing problem. For a set of preselected customers, we are given the shortest and longest possible service time of each customer as well as the total number of no-shows. We limit each customer's waiting time, even in a worst-case scenario, to a predetermined tolerance for waiting (so-called maximum waiting time guarantee). Then we minimize a weighted sum of idle times and overtime (so-called total cost).
The main contributions of this work can be summarized as follows.

\textbf{Maximum Waiting Time Guarantees.} To the best of our knowledge, this paper is the first to introduce maximum waiting time guarantees for intraday appointment scheduling. The guarantees ensure that each customer's waiting time stays below a given individual threshold, even in a worst-case scenario. Hence, they ensure that all customers experience acceptable waiting times and prevent customer dissatisfaction due to excessive waiting times. In addition, communicating these guarantees can increase customer satisfaction \citep{kumar1997impact}.

\textbf{Optimal Scheduling and Sequencing Rules.} We show that the robust appointment scheduling and sequencing problem with customer-specific maximum waiting time guarantees is $\mathcal{NP}$-hard in general and introduce a mixed-integer linear programming formulation. We present optimal solutions for polynomial-time solvable special cases: For constant idle time costs and zero no-shows, we prove that it is optimal to sort customers in ascending order of the uncertainty in their service times and their waiting time guarantees, resembling the well-known Smallest-Variance-First rule. For non-increasing idle time costs and any given but fixed customer sequence, we prove optimality of the so-called ASAP rule, which schedules customers as early as possible while respecting their waiting time guarantees.

\textbf{Computational Study Using Patient Data from a Radiology Department.}
Using real data from the radiology department of a large hospital, we first demonstrate that the problem can be solved within reasonable computation times for all considered instances. Second, we conduct a sensitivity analysis of the waiting times, idle times, and overtime obtained with optimal sequences and schedules. Third, we compare these results with those obtained using a weighted sum approach. The findings indicate that incorporating waiting time guarantees results in limited and more balanced waiting times across all customers. Moreover, our approach introduces these guarantees with a minimal increase in the worst-case total cost. In particular, in instances where the waiting times obtained with the weighted sum approach are also acceptable for all patients, our approach generally yields lower total costs.

The remainder of the paper is organized as follows. Section~\ref{subSection-literature} reviews related optimization problems in the appointment scheduling literature. Section~\ref{Section-model} formally introduces the Robust Appointment Scheduling Problem with Waiting Time Guarantees. We show that it is $\mathcal{NP}$-hard in general and derive a mixed-integer linear program formulation. In Section~\ref{subsection-rules}, we present optimal scheduling and sequencing rules for relevant special cases in polynomial time. In Section~\ref{subsection-experiments}, we conduct a computational study with real-world data, showing that introducing maximum waiting time guarantees can improve both waiting times and total cost compared to a weighted sum approach.

\section{Literature Review}\label{problemdescription} 
\label{subSection-literature}

For a general overview of appointment scheduling, we refer the interested reader to \cite{cayirli2003outpatient, gupta2008appointment, ahmadi2017outpatient}, and \cite{marynissen2019literature}. In the following, we review two streams of the appointment scheduling literature that are most relevant to our research. In Section \ref{RASreview}, we discuss robust optimization approaches using box uncertainty sets for service times. In Section \ref{WAITreview}, we survey alternative approaches for modeling waiting time, beyond the widely used weighted sum approach that minimizes total idle time, total waiting time, and overtime.

\subsection{Robust Optimization Framework} \label{RASreview}
Uncertainty in appointment scheduling is typically addressed using one of three main frameworks: stochastic programming, distributionally robust optimization, or robust optimization. Assuming full knowledge of the probability distributions of uncertain parameters, stochastic programming typically seeks to optimize expected performance. Both single-stage and two-stage stochastic programming models have been widely applied to appointment scheduling problems (\cite{cayirli2003outpatient, gupta2008appointment, ahmadi2017outpatient}). When true probability distributions are not known with certainty, distributionally robust optimization provides an alternative by optimizing expected performance under the worst-case distribution within an ambiguity set. While being computationally challenging, they have been successfully applied to appointment scheduling (\cite{kong2013scheduling, mak2015appointment, zhang2017distributionally, jiang2017integer, jiang2019data, kong2020appointment}). Finally, when individual customer experiences are not accurately represented by expected performance or other probabilistic measures, robust optimization provides a compelling alternative by mitigating unfavorable singular outcomes through worst-case performance optimization over predefined uncertainty sets—such as box, polyhedral, ellipsoidal, or general convex sets. Foundational contributions in this area include those by \cite{el1997robust, el1998robust, ben1998robust, bertsimas2004price, ben2000robust}, and \cite{ben2009robust}, with a comprehensive overview provided by \cite{lu2021review}.

Robust optimization approaches to appointment scheduling are scarce. To the best of our knowledge, \cite{mittal2014robust} present the first such approach, minimizing a weighted sum of idle time costs and waiting time costs under box uncertainty sets for service times. \cite{schulz2019robust} generalize the problem in \cite{mittal2014robust} to idle time costs that are heterogeneous in the customer sequence and thus implicitly heterogeneous over time. Both works derive scheduling and sequencing rules for various special cases that are remarkably easy to implement. Given a customer sequence and homogeneous idle time costs, \cite{mittal2014robust} present a polynomial-time algorithm to compute optimal interappointment times, which decrease as the planning horizon progresses. They also show that a simple variant of Smith's rule, scheduling customers in ascending order of the length of their service time intervals divided by their waiting time costs, is a 1.137 approximation. \cite{schulz2019robust} prove a slightly adapted variant of this rule to be 1.0604-approximate, and the problem of simultaneously scheduling and sequencing customers with heterogeneous idle time costs and the weighted sum objective function to be strongly $\mathcal{NP}$-hard. \cite{issabakhsh2021scheduling} address a robust appointment scheduling problem for infusion centers and solved it with a scheduling heuristic based on Adaptive Large Neighborhood Search. 
\cite{brugmanrobust} recently introduced a robust optimization approach for appointment scheduling that minimizes the sum of worst-case costs from idle time, waiting time, and overtime over general convex uncertainty sets, and presented the first exact solution method for this problem class. Notably, the approach of \cite{brugmanrobust} may offer a foundation for extending our work to more general uncertainty sets. While our analysis focuses on uncertain service times modeled via box uncertainty sets, chosen for their analytical tractability and ease of implementation, our contribution further adds a new perspective to this literature stream by incorporating individual waiting time guarantees. 

\subsection{Modeling Waiting Time} \label{WAITreview}
Building on the classification by \cite{bertsimas2011price}, we divide approaches for modeling waiting times into two substreams, based on the criterion used to evaluate waiting time. The utilitarian criterion minimizes a weighted sum of waiting times, while fairness criteria address individual waiting times more directly.

\subsubsection{Utilitarian Criterion.} In the first substream, we review appointment scheduling approaches that deviate from the classic weighted sum objective function but retain the utilitarian criterion. 
\cite{sang2021appointment} use a stochastic optimization approach to minimize any quantile of the total (idle time, overtime, and waiting) cost for a given job sequence. They find a semi-dome-shaped pattern with increasing interappointment times at the beginning of the planning horizon and constant interappointment times later. \cite{zhou2019intraday} use a chance-constraint approach, minimizing, among others, the probability that the sum of waiting times exceeds a certain threshold. They find optimal interappointment times to follow the well-known dome-shaped pattern in the case of zero no-shows and walk-ins. With a positive number of walk-ins, they find optimal interappointment times in the middle of the planning horizon to be constant, forming a so-called plateau dome-shaped pattern.  

\subsubsection{Fairness Criteria.} 
\cite{millhiser2012assessing} propose a paradigm shift by evaluating appointment schedules based on probabilities of exceeding waiting time, idle time, and overtime thresholds, rather than optimizing a fixed cost function. 
\cite{wang2014service} characterize individual waiting time distributions and derive performance measures such as expected waiting and completion times. \cite{millhiser2015designing} design schedules where each patient’s probability of waiting longer than a target threshold is uniform across all patients. 
\cite{qi2017mitigating} lexicographically minimizes the so-called delay unpleasantness of the longest-waiting customer, which considers the frequency and intensity of delays above a certain threshold. Optimal scheduling rules depend on waiting time thresholds and the weighting of overtime costs. As the weighting of overtime increases, the interappointment times between the first appointments are shorter than with weighted sum approaches. As the waiting time threshold decreases, later start times are assigned to the later appointments in the planning horizon. Following a queuing approach, \cite{benjaafar2020appointment}, by contrast, assign the earliest possible appointment time to each customer request while limiting the expected waiting time of each customer. They find that the optimal times between appointments increase in the order of arrival. Similarly, \cite{jouini2022appointment} derive an upper bound on expected waiting times with a queueing-based model that accounts for random service times, non-punctuality, and no-shows. Finally, for a given customer sequence, \cite{bendavid2018developing} minimize the completion time of the last appointment while bounding the probability of delay for each appointment. Under the assumption that the distribution of service times is known, they develop a simulation-based sequential algorithm and show that optimal interappointment times follow a dome-shaped pattern.

\subsubsection{Our Contribution.} In this paper, we contribute to fair appointment scheduling by providing an a priori waiting time guarantee for each customer, effectively limiting their worst-case waiting time. These waiting time guarantees can be customized to match the tolerance levels for waiting identified in empirical studies.
We derive optimal closed-form scheduling and sequencing rules. Moreover, compared to existing approaches, we allow for substantial cost reduction by minimizing (heterogeneous) cost incurred only by idle time and overtime, accounting for no-shows, and the sequencing decision.

\section{Robust Appointment Scheduling with Waiting Time Guarantees}
\label{Section-model}

This section is organized as follows. In Section~\ref{mmodel}, we formally introduce the Robust Appointment Scheduling Problem with Waiting Time Guarantees. In Section~\ref{NPsection}, we show that the problem is $\mathcal{NP}$-hard. In Section~\ref{SectionMILP}, we derive two mixed-integer linear programs, the first for the case that all customers show up and the second for the general case, which considers no-shows.

\subsection{Mathematical Model} \label{mmodel}

We consider a set of $n$ preselected customers denoted by $[n] := \lbrace 1, \ldots, n\rbrace$, and a single service provider with a regular working time of $L$ time units. Our decisions involve determining the customer sequence and appointment times. To determine the customer sequence, we introduce binary variables $\pi_{ij}$ for $i,j \in [n]$, where $\pi_{ij} = 1$ if and only if exactly $i-1$ customers are served before customer $j$. In other words, customer $j$ is assigned to the $i$-th appointment. We define continuous variables $A_i \geq 0$ to represent the start time of each appointment $i \in [n]$. We consider the following costs.
Idle time before the first appointment incurs a per-time unit cost $c_1$, and idle time between two consecutive appointments $i-1$ and $i$ incurs a per-time unit cost $c_i$, for $2 \leq i \leq n$. If there is any idle time before the planning horizon of $L$ time units ends, it incurs a per-time unit cost $c_{n+1}$. Additionally, overtime incurs a per-time unit cost~$c_o$. We aim to determine a customer sequence and appointment times that minimize the cost incurred by idle time and overtime, while ensuring that no customer $j \in [n]$ waits longer than their maximum waiting time guarantee $W_j$.

The main difficulty is that customers have uncertain service times and no-show behavior.
For each customer $j \in [n]$, we only have information about the shortest and longest possible service times, denoted as $\ubar{p}_j$ and $\bar{p}_j$, respectively. This leads to the scenario set of possible service time realizations $\mathcal{P} := [{\ubar{p}_1}, {\bar{p}_1}] \times \ldots \times [{\ubar{p}_n}, {\bar{p}_n}]$. Additionally, we know the total number of customers who will show up within the planning horizon, denoted as $k$, where $k \leq n$. We have no information on which $k$ out of $n$ customers show up. This leads to the scenario set of possible no-show realizations $\Lambda_k := \lbrace {\lambda} \in \lbrace 0, 1 \rbrace^n : \sum_{j = 1}^n {\lambda_j} = k \rbrace$, where ${\lambda_j} = 1$ if and only if customer $j$ shows up. We assume that customers who show up are on time.

For any given but fixed service time scenario ${ p} \in \mathcal{P}$ and no-show scenario ${  \lambda} \in \Lambda_k$, and given $\pi$ and A, we compute waiting times, idle times, and overtime as follows. We start by initializing $C_0(\pi, A, { p}, {\lambda}) : = 0$, and define completion times recursively for each appointment $i \in [n]$ as ${C}_{i}(\pi, A, { p}, {\lambda}) := \max\left\lbrace {A}_{i}, {C}_{i-1}(\pi, A, { p}, {\lambda})\right\rbrace + \sum_{j=1}^n \pi_{ij}{{\lambda}_{j}{p}_{j}}$. It is easy and useful to see by induction that the completion times can be reformulated as $C_i(\pi, A, { p}, {\lambda}) = \max_{l \in [1,i]} \left\{{A}_{l} + \sum_{k = l}^{i} \sum\limits_{j = 1}^n \pi_{kj}{{\lambda}_{j}{p}_{j}} \right \}$, see Lemma~\ref{completiontime} in the \hyperref[AppLit]{Online Appendix}. With $(\cdot)^+ := \max\{0, \cdot\}$, the waiting time for each appointment $i \in [n]$ is calculated as the time between the completion of the previous appointment and the start of the current appointment if the scheduled customer shows up; otherwise, it is 0:
$$\textrm{wait}_i\textcolor{black}{(\pi, A, { p}, {\lambda})} := \sum_{j=1}^n \pi_{ij} {\lambda_j} \cdot  \left( C_{i-1}\textcolor{black}{(\pi, A, { p}, { \lambda})} - A_i \right)^+ $$ Similarly, the idle time before each appointment $i \in [n+1]$, with $A_{n+1} := L$, is given by $\left( A_i - C_{i-1}\textcolor{black}{(\pi, A, { p}, {\lambda})}\right)^+$, and overtime is computed as $(C_n\textcolor{black}{(\pi, A, { p}, {\lambda})} - L)^+$.
The cost function
$$
\textrm{cost}\textcolor{black}{(\pi, A, { p}, { \lambda})} := \sum\limits_{i = 1}^{n} c_i \left( {A}_i - {C}_{i-1}\textcolor{black}{(\pi, A, { p}, {\lambda})} \right)^+ + \max\left\lbrace c_{n+1}(L - C_n\textcolor{black}{(\pi, A, { p}, {\lambda})}), c_o( C_n\textcolor{black}{(\pi, A, { p}, {\lambda})} - L) \right\rbrace
$$ accounts for the total cost arising from idle time and overtime. We refer to Table~\ref{tab:notation} in the Online Appendix for a summary of the notation.

To establish sustainable solutions, appointment schedules should be designed sufficiently robust to withstand unfortunate service times and no-shows. \emph{Excessive} waiting times, idle time, and overtime not only incur high costs but also lead to employee and customer stress, dissatisfaction, and reduced loyalty. In healthcare, these factors can even harm patient health \citep{bae2014assessing}. Hence, we introduce a robust optimization approach that ensures smooth operations for all stakeholders, even in adverse scenarios. This gives rise to the following Robust Appointment Scheduling Problem with Waiting Time Guarantees \textcolor{black}(RASWTG):
\begin{align} \label{RASWTG}
\!\min_{\substack{\pi,A}} \,& \, U & \\ \label{rcon}
\text{s.t.}\, & \max_{\substack{{\boldsymbol p} \in \mathcal{P}, {\boldsymbol \lambda} \in \Lambda_{k} }} \left\lbrace \textcolor{black}{\max}\left\lbrace \textrm{cost}\textcolor{black}{( \pi, A, {\boldsymbol p}, {\boldsymbol\lambda})}
- U , \max\limits_{i \in [n]} \left\{ \textrm{wait}_i\textcolor{black}{(\pi, A, {\boldsymbol p}, {\boldsymbol\lambda})} - \sum_{j=1}^n \pi_{ij} W_j \right\} \right\rbrace \right\rbrace \leq 0 & \\
& \sum_{i=1}^n \pi_{ij} = 1, \sum_{j=1}^n \pi_{ij} = 1 \qquad \forall i,j \in [n] & \\
& A_1 = 0 \\
& \pi_{ij} \in \lbrace0,1\rbrace \qquad \forall i,j \in [n] &
\end{align}
The {Objective Function}~\hyperref[RASWTG]{(1)} minimizes the upper bound $U$ on total cost incurred by idle time and overtime. Constraint~\hyperref[RASWTG]{(2)} enforces this upper bound to remain valid even in worst-case scenarios, preventing excessive idle time and excessive overtime. Additionally, it ensures that the waiting time guarantees of all customers are met across all scenarios, ensuring acceptable waiting times. Thus, by preparing both the provider and the customers for their respective worst-case scenarios, Constraint~\hyperref[RASWTG]{(2)} ensures smooth operation even on a challenging day. 
Constraints~\hyperref[RASWTG]{(3)} enforce that each customer is assigned to exactly one appointment, and each appointment is assigned to exactly one customer. Constraint~\hyperref[RASWTG]{(4)} sets the start time of the first appointment to $0$.

\begin{remark}
Due to the adversarial problems in Constraint~\hyperref[RASWTG]{(2)}, which maximize the total cost and the waiting time of the longest-waiting customer, the above problem formulation is non-linear. In particular, it is not amenable to standard MIP solvers in this form.
\end{remark}

\begin{remark}
The problem always has a feasible solution. For an arbitrary but fixed sequence $\pi$, it is easy to see that the appointment times $A_1 := 0$ and $A_i := \left( \max_{l \in [1,i-1]} \left( {A}_{l} + \sum_{k = l}^{i-1} \sum_{j = 1}^n \pi_{kj} {\bar{p}_{j}} \right) - \sum_{j=1}^n \pi_{ij}W_j \right)^+$ for $i \in [2,n]$ are feasible solutions.
\end{remark}

\begin{remark}
It is noteworthy that a monotonic increase in the appointment times is not guaranteed \textcolor{black}{(see Appendix~\ref{proof00o}, page~\pageref{proof00o})}. 
 \label{remarkwithproof}
\end{remark}

\tikzstyle{block} = [rectangle, draw, text width = 16em, text centered, node distance=1cm, minimum height=1.5em, minimum width = 2.5cm]
\tikzstyle{decision} = [rectangle, draw, text width = 25em, text centered, node distance=1cm, minimum height=1.5em, minimum width = 5cm]
\tikzstyle{line} = [draw, -latex']
\tikzstyle{cloud} = [draw, ellipse, node distance=2.75cm, minimum height=5em, text width = 6em, text centered]
\tikzstyle{cloud1} = [draw, ellipse, node distance=2.75cm, minimum height=5em, text width = 6em, text centered]

\subsection[Complexity Analysis]{Complexity Analysis}\label{NPsection}
In this section, we establish the $\mathcal{NP}$-hardness of the Robust Appointment Scheduling Problem with Waiting Time Guarantees, which holds even under the assumption of zero no-shows and identical waiting time guarantees for all customers. This result expands upon existing $\mathcal{NP}$-hardness proofs for appointment scheduling and sequencing problems employing weighted sum approaches \citep{bosch1997scheduling, kong2016appointment, schulz2019robust}. 

\begin{theorem}[$\mathcal{NP}$-Hardness]\label{NP}
Let all customers show up and have identical waiting time guarantees $W$. Moreover, let the idle time costs fulfill $c_1 = \ldots = c_{\sigma} > c_{\sigma +1} = \ldots = c_n$ for some $\sigma \in [n-1]$ and the overtime cost $c_o = 0$. Then the Robust Appointment Scheduling Problem with Waiting Time Guarantees is $\mathcal{NP}$-hard.
\end{theorem}

We briefly sketch the proof and refer to the \hyperref[NPproof]{Online Appendix} for details. In a worst-case scenario, worst-case idle times arise in two ways: If no idle time occurs up to and including appointment $i ^* -1 $, then the idle time before appointment $i^*$ equals $\sum_{i=1}^{i^*-1}\sum_{j=1}^{n}\pi_{ij}( {\bar{p}_j}-{\ubar{p}_j}) - W$. The idle time before all subsequent appointments $i > i^*$ equals $\sum_{j=1}^{n}\pi_{i-1,j}( {\bar{p}_j}-{\ubar{p}_j})$. In the case of $i^* = \sigma$, the unit idle time cost before appointment $i^*$ exceeds all later unit costs of idle time. The challenge lies in partitioning the customers into two groups such that the idle time sum $\sum_{i=1}^{i^*-1}\sum_{j=1}^{n}\pi_{ij}( {\bar{p}_j}-{\ubar{p}_j}) - W$ is minimized. In the \hyperref[NPproof]{Online Appendix}, we show that the well-known Subset Sum Problem can be reduced to this particular problem, establishing the $\mathcal{NP}$-hardness of \hyperref[RASWTG]{RASWTG}.

\begin{remark}
If the assumption on idle time costs in Theorem~\ref{NP} is replaced by constant idle time costs $c_1 = \ldots = c_n$, indicating no time preference for idle time occurrence, the Robust Appointment Scheduling Problem with Waiting Time Guarantees becomes solvable in polynomial time. We discuss this case in detail in Section~\ref{subsection-rules}.
\end{remark}

\subsection[Linearization]{Linearization} \label{SectionMILP}
Due to the recursive nature of completion times in the cost function, applying the classical dualization scheme by \cite{bertsimas2004price} for linearization is impractical. However, we overcome this challenge by exploring structural problem properties. In Section~\ref{MIP0}, we leverage structural properties to determine worst-case service times analytically. This allows us to formulate a compact Mixed-Integer Linear Program (MILP) for the case of zero no-shows. Section~\ref{ad-noshows} then motivates the consideration of no-shows and outlines advanced structural properties utilized for linearization in the presence of no-shows. For a detailed consideration of no-shows as well as for proofs not included in the main paper, we refer the interested reader to the \hyperref[AppLit]{Online Appendix}~\ref{A_MILP}.

\subsubsection{Worst-Case Service Time Scenarios.}\label{MIP0}

We investigate service times in adversarial scenarios maximizing cost and waiting times, respectively. Consequently, we present a compact MILP for the case of zero no-shows.

\paragraph{Adversarial Problem of Maximizing Cost.}
The following theorem shows that the total cost is maximized in one of the $n + 1$ sequence-dependent scenarios, in which customers scheduled for the first $i^* \in \lbrace 0, \ldots, n \rbrace$ appointments require minimum service duration (idle time maximization), and all subsequent customers require maximum service time (overtime maximization).
\begin{theorem} \label{wc} Given a customer sequence $\pi$, appointment times $A$, and ${\lambda} \in \Lambda_k$, let $\ubar{C}_{i}(\pi, A, {\lambda}) := \max\left\lbrace {A}_{i}, \ubar{C}_{i-1}(\pi, A, {\lambda})\right\rbrace + \sum_{j=1}^n \pi_{ij}{{\lambda}}_{j}{\ubar{p}_{j}}$ for $i \in [n]$, with $\ubar{C}_0(\pi, A, {\lambda}) := 0$, define the completion times when all customers require minimum service time. Then $ \max_{{\boldsymbol p} \in \mathcal{P}} \textrm{cost}\textcolor{black}{({\pi, A, \boldsymbol{p}, \lambda})} $ is equivalent to $$ \max_{i^* \in \lbrace0, \ldots, n+1\rbrace} \sum_{i=1}^{i^*} c_i(A_i - \ubar{C}_{i-1}(\pi, A, {\lambda}))^+ + c_o\left( A_{i^*} + \sum_{k = i^*}^n \sum_{j = 1}^n \pi_{kj}{\lambda_j}\bar{p}_j - L \right)^+, $$ where $A_{n + 1} := L$.
\end{theorem}

\paragraph{Adversarial Problem of Maximizing Waiting Times.} 
As for each no-show scenario ${\lambda} \in \Lambda_k$, service time scenario $p \in \mathcal{P}$, and patient $j \in [n]$, the realized service time ${\lambda_{j}}{p}_{j}$ is non-negative, it is straightforward to see that the scenario of maximum service times \begin{align} \label{pmax}
{ \bar{p}} :=({\bar{p}_1}, \ldots, {\bar{p}_n}) \in \argmax_{\substack{{\boldsymbol p} \in \mathcal{P} }} \left\lbrace \max_{l \in [1,i-1]} \left ({A}_{l} + \sum_{k = l}^{i-1} \sum_{j = 1}^n \pi_{kj} { {\lambda}_{j}\boldsymbol{p}_{j}} \right ) 
\right\rbrace, \textmd{}\end{align}
uniformly maximizes all completion times and therefore the waiting time for each appointment $i \in [n]$ in Constraint~\hyperref[RASWTG]{(2)}.

\paragraph{MILP.} Now, let $\Delta_i$ denote the idle time before appointment $i \in [n + 1]$ and $\sigma_{i^*}$ denote the overtime accumulated starting from appointment $i^* \in [n + 1]$. Then the worst-case service time scenarios from Theorem~\ref{wc} allow us to obtain the following mixed-integer linear program for the case of zero no-shows (RASWTG-$0$):
\begin{align} \label{RASWTG-n}
\min \quad & U & \\
\text{s.t.} \quad
& U \geq \sum_{i = 1}^{i^*} c_i \Delta_i + c_o \sigma_{i^*} & \forall {i^*} \in [ n+1] \\
& \Delta_i \geq {A}_i - \ubar{C}_{i-1} & \forall i \in [n+1] \\
& \ubar{C}_i = \max\left\lbrace A_i, \ubar{C}_{i-1} \right\rbrace + \sum_{j=1}^n \pi_{ij}\ubar{p}_{j} & \, \forall i \in [n] \\ 
& \sigma_{i^*} \geq A_{i^*} + \sum_{k = i^*}^n\sum_{j=1}^n \pi_{kj}\bar{p}_j - L & \forall {i^*} \in [n] \\
& A_{l} + \sum\limits_{k = l}^{i-1} \sum_{j = 1}^n \pi_{kj} \bar{p}_{j} - A_{i} \leq \sum_{j=1}^n \pi_{ij}W_j & \forall i, l: i \in [n] , l \in[i-1] \\
& \sum_{j=1}^n \pi_{ij} = 1, \sum_{i=1}^n \pi_{ij} = 1 & \forall i,j \in [n] \\
& A_1 = 0, \ubar{C}_0 = 0, A_{n+1} = L & \label{2-0ns} \\
& A_i, \ubar{C}_i \geq 0 & \forall {i} \in [n] \\
& \Delta_i, \sigma_{i} \geq 0 & \forall {i} \in [n+1] \\
& \pi_{ij} \in \lbrace 0,1 \rbrace & \forall i,j \in [n]
\end{align}
Compared to \hyperref[RASWTG]{(1 - 5)}, \hyperref[RASWTG-n]{RASWTG-0} can be characterized as follows: Based on Lemma~\ref{wc}, Constraints~\hyperref[RASWTG-n]{(8) - (11)} determine the worst-case total cost incurred by idle time and overtime. Based on (\ref{pmax}), Constraint~\hyperref[RASWTG-n]{(12)} ensures that the waiting time guarantees are fulfilled even in the worst case.

\hyperref[RASWTG-n]{RASWTG-$0$} is an assignment problem of customers to appointments with assignment-dependent cost. It has $O(n^2)$ binary variables, $O(n)$ continuous variables, and $O(n^2)$ constraints. Using standard linearization techniques, Constraint~\hyperref[RASWTG-n]{(10)} can be linearized by introducing another $O(n)$ binary variables, see Remark \ref{MIPrefs} in the \hyperref[AppLit]{Online Appendix}.
The computational study in Section~\ref{subsection-experiments} shows that \hyperref[RASWTG-n]{RASWTG-$0$} is solvable with a state-of-the-art MIP solver in short times for all instances with up to twenty patients from a data set of CT examinations in a large university hospital.

\subsubsection{Worst-Case No-Show Scenarios.}\label{ad-noshows}

No-shows introduce uncertainties that, if overlooked, can strongly elevate costs. 
We now address the task of formulating a MILP for the general Robust Appointment Scheduling Problem with Waiting Time Guarantees and $k$ no-shows, $0 \leq k \leq n$. To this end, we explore the adversarial problems, seeking no-show scenarios that maximize waiting times and total cost, respectively. Identifying worst-case no-show scenarios is notably more complex than determining worst-case service times. Moreover, maximizing total cost is particularly intricate because the cost incurred by a no-show is not only proportional to the customer's service time but also encompasses sequence-dependent idle time and overtime costs. For a comprehensive understanding of the technical aspects, we direct the interested reader to the \hyperref[AppLit]{Online Appendix}~\ref{A_MILP}. Here, we summarize our key findings:

\paragraph{Adversarial Problem of Maximizing Cost.} In Lemma~\ref{wcc} in the \hyperref[AppLit]{Online Appendix}, we characterize a worst-case no-show scenario for an arbitrary but fixed sequence. Specifically, it suffices to consider, for each $i^* \in \lbrace 1, \ldots, n + 1 \rbrace$, a no-show scenario in which, among the customers scheduled for the last $n - i^* + 1$ appointments, those with the longest maximum service times show up. If $k > n - i^* + 1$, all customers scheduled for these last appointments show up. Additionally, among the customers scheduled for the first $i^* -1 $ appointments, the $k - (n - i^* + 1)$ ones with the smallest product of idle time cost and minimum service time show up.

\paragraph{Adversarial Problem of Maximizing Waiting Times.} 
The following lemma establishes the equivalence of (i) computing a single worst-case scenario that maximizes the waiting time for the longest waiting customer and (ii) independently computing a worst-case scenario for each customer that maximizes their waiting time. We note that it suffices to consider no-show scenarios $\lambda \in \Lambda_{k-1}$, as waiting time is incurred only if the customer scheduled for the $i$-th appointment shows up. This inherent decomposability of the adversary problem for waiting times stems from the scenario-independent appointment times.

\begin{lemma}[Decomposability] \label{sep} Given a customer sequence $(\pi_{ij})_{i,j \in [n]}$ and appointment times $(A_i)_{i \in [n]}$, it holds that
$\max_{\boldsymbol{p} \in \mathcal{P}, \boldsymbol{\lambda} \in \Lambda_{k-1}} \left\lbrace \max_{i \in [n]} \left\{ C_{i-1}(\pi, A, \boldsymbol{p}, \boldsymbol{\lambda}) - A_i - \sum_{j=1}^n \pi_{ij} W_j \right\} \right\rbrace \leq 0 $ if and only if $ \max_{i \in [n]} \left\lbrace \max_{\boldsymbol{p^{(i)}} \in \mathcal{P}, \boldsymbol{\lambda^{(i)}} \in \Lambda_{k-1}} \left\{ C_{i-1}(\pi, A, \boldsymbol{p^{(i)}}, \boldsymbol{\lambda^{(i)}})- A_i - \sum_{j=1}^n \pi_{ij} W_j \right\} \right\rbrace \leq 0$.
\end{lemma}

By employing the completion time reformulation, leveraging the total unimodularity of the constraint matrix associated with the adversarial problem for each appointment, and applying LP duality, the adversarial problem of maximizing waiting times is then reduced to $\mathcal{O}(n)$ linear programming problems. Each of these problems entails $\mathcal{O}(n)$ variables and constraints.


\paragraph{MILP.}
These results lead to a comprehensive MILP for the general case, detailed in the \hyperref[AppLit]{Online Appendix}~\ref{A_MILP} and denoted as \hyperref[RASWTG-k]{RASWTG}. This MILP encompasses $O(n^3)$ variables and constraints, where the constraint matrix exhibits sparsity. The computational study in Section~\ref{subsection-experiments} demonstrates the efficient solvability of \hyperref[RASWTG-k]{RASWTG} for all instances involving up to ten patients from a dataset of CT examinations in a large hospital, employing a state-of-the-art MIP solver. \textcolor{black}{Future work could explore the tightness of continuous relaxations as well as decomposition methods.}

\section{Optimal Scheduling and Sequencing }
\label{subsection-rules}

In this section, we present optimal scheduling and sequencing rules to solve \hyperref[RASWTG]{RASWTG} in polynomial time in special cases, as summarized in Table~\ref{tab:methods}. Additionally, we investigate optimal interappointment times for the general case. All results demonstrate how to provide maximum waiting time guarantees cost-efficiently.

\begin{table}[h]
\centering \small
\caption{Problem Variants and Solution Methods}
\begin{tabular}{lclccclccccccccccccccccccccccccccc} \toprule
&& \bf Problem Variant && && \bf Solution Method \\ \midrule
Section \ref{seq} && Constant idle time cost \& zero no-shows && && \hyperref[opt-sequence-general]{Polynomial-time sequencing rule} \\
Section \ref{sched} && Non-increasing idle time cost \& fixed sequence && && \hyperref[algcap]{Polynomial-time scheduling rule} \\
Section \ref{interapp} && General case && && \hyperref[RASWTG-k]{MILP} \\
\bottomrule
\end{tabular}
\label{tab:methods}
\end{table}

\subsection{Polynomial-Time Sequencing for the Case of Constant Idle Time Costs and Zero No-Shows} \label{seq}
Finding an optimal customer sequence can substantially reduce idle time costs \citep{cayirli2008assessment, deceuninck2018outpatient} but appears extremely difficult \citep{mak2015appointment}. Similarly, Theorem~\ref{NP} implies that (unless $\mathcal{P} = \mathcal{NP}$) we cannot find a universal optimal scheduling and sequencing rule for the Robust Appointment Scheduling Problem with Waiting Time Guarantees in polynomial time. However, for a relevant special case, we now prove optimality of a polynomial-time sequencing rule consistent with the well-known Smallest-Variance-First (SVF) heuristic.

SVF is the most common heuristic for weighted sum approaches to appointment sequencing with stochastic service times. It ranks customers in ascending order of variance, following the intuition to reduce
the probability of deviating from the schedule by placing customers with less variable service times first. For recent SVF studies in a stochastic optimization framework, we refer to \cite{kong2016appointment, jafarnia2020non}, and \cite{ de2021performance}. Moreover, \cite{mak2015appointment} prove, under a mild condition, the optimality of the SVF rule for a robust mean-variance model with an arbitrary number of appointments. Recently, \cite{kong2016appointment} showed that the optimality of the SVF policy depends on the number of customers as well as the shape of service time distributions.

\begin{lemma}[Optimal Customer Sequence] \label{opt-sequence-general} Let the number of no-shows equal zero and idle time costs $c_1 = \ldots = c_{n+1}$ be constant, indicating no time preference for the occurrence of idle time. Then it is optimal to schedule customers $j \in [n]$ in non-decreasing order w.r.t. $\bar{p}_j - \ubar{p}_j + (1+c_o)W_j$.
\end{lemma}

Under the given assumptions, it is optimal to schedule customers $j \in [n]$ in non-decreasing order with respect to their service time uncertainty $\bar{p}_j - \ubar{p}_j$ and their waiting time guarantee $W_j$, as a function of overtime cost $c_o$. This result intuitively extends the SVF rule to account for waiting time guarantees:
Uncertainty is kept as low as possible for as long as possible, particularly for customers with stricter waiting time guarantees. The relative importance of both criteria depends on the overtime cost. We call it the Smallest-Variance-First Rule with Waiting Time Guarantees (SVF-WTG). We will further determine optimal appointment times under these (and milder) assumptions in Section~\ref{sched}.

\subsection{Polynomial-Time Scheduling Rule under Non-Increasing Idle Time Costs} \label{sched}

In the following, we assume that an arbitrary sequence of customers is given. We first present the so-called As~Soon~As~Possible (ASAP) scheduling rule defined by Algorithm~\ref{algcap}. Second, we show that in the case of non-increasing idle time costs, it leads to an optimal solution in polynomial time. Third, we describe how ASAP generalizes the Bailey-Welch rule \citep{welch1952appointment}.

\begin{algorithm}[b]
\caption{ASAP Scheduling Rule}\label{algcap}
\begin{algorithmic}[1]
\Require $n \in \mathds{N}, c \in \mathds{R}^{n+1}$ with $c_1 \geq \ldots \geq c_{n+1}$, $(W_j)_{j \in[n]}$, $\bar{p} \in \mathds{R}^n$, $k \geq 0$, $\pi \in \lbrace 0,1 \rbrace^{n \times n}$
\State $A_1 \gets 0$
\For{$i \gets 2$ to $n$}
\For{$l \gets 1$ to $i-1$}
\If{$i-l>k-1$}
\State $U_l \gets \min_{\alpha, z} \left\lbrace (k-1)\,\alpha + \sum_{\sigma=l}^{i-1} z_{\sigma} \,:\, z_{\sigma} + \alpha \geq \sum_{j = 1}^n \pi_{\sigma j} \bar{p}_{j}, \, z_{\sigma} \geq 0 \,\, \forall {\sigma} \in [l,i-1] \right\rbrace$
\Else \State $U_l \gets \sum_{\sigma = l}^{i-1} \sum_{j = 1}^n \pi_{\sigma j} \bar{p}_{j}$
\EndIf
\EndFor
\State $A_{i} \gets \left( \max_{l \in [i-1]} A_{l} + U_l- \sum\limits_{j=1}^n \pi_{ij} W_{j} \right)^+$

\EndFor
\State \Return $(A_1, \ldots, A_n)$
\end{algorithmic}
\end{algorithm}

The ASAP scheduling rule schedules each customer as early as possible while respecting the waiting time constraints. It is constructed as follows by Algorithm~\ref{algcap}. The start time of the first appointment is set to $0$ in Line~\hyperref[algcap]{1}. For each subsequent appointment, the algorithm computes the latest possible completion time of the previous appointment using Lines~\hyperref[algcap]{5} and \hyperref[algcap]{7}. \textcolor{black}{
Line~\hyperref[algcap]{5} calculates the upper bound $U_l$ of service times between appointments $l$ (inclusive) and $i$ (exclusive). Specifically, if the number of no-shows $k$ (minus 1 because the customer for appointment $i$ is assumed to show up) exceeds $i - l$ (as in Line~\hyperref[algcap]{7}), then the upper bound is achieved when all customers show up. Otherwise, it corresponds to the sum of the longest service times among those who do show up. }
In the special case of zero no-shows, Lines~\hyperref[algcap]{3} to~\hyperref[algcap]{7} can be skipped, and Line~\hyperref[algcap]{8} can be replaced by
$
A_{i} = \left( \,\sum_{k=1}^{i-1} \sum_{j=1}^n \pi_{kj} \bar{p}_{j}- \sum_{j=1}^n \pi_{ij}W_{j}\right)^+$.

Algorithm~\ref{algcap} computes optimal appointment times for a given sequence under non-increasing idle time costs, as stated in the following theorem. This condition, while crucial for the optimality proof (see Remark~\ref{remark_inc} in the \hyperref[theoremproof]{Online Appendix}), holds practical significance.
Constant idle time costs indicate no time preference for the occurrence of idle time. Decreasing idle time costs indicate that later idle time is less expensive than earlier idle time, potentially due to its productive use for documentation work incurred by earlier appointments.

\begin{theorem}[Optimal Appointment Times] \label{opt-times} Given non-increasing idle time costs $c_1 \geq \ldots \geq c_{n+1}$, and a customer sequence $\pi$, Algorithm~\ref{algcap} computes optimal appointment times in polynomial time for the Robust Appointment Scheduling Problem with Waiting Time Guarantees.
\end{theorem}

The ASAP rule can be considered a generalization of the well-known Bailey-Welch rule, which schedules the first two customers simultaneously and all subsequent customers sequentially. In Line~\hyperref[algcap]{8}, each customer is scheduled to arrive at time $0$ as long as the latest possible completion times of the previous customers are not longer than their waiting time guarantee. If this condition is no longer met, customers are scheduled sequentially. In particular, the maximum waiting time guarantees control how many customers can be scheduled at the start time of the planning horizon.

\subsection{Optimal Interappointment Times for Varying Costs and No-Show Rates}\label{interapp} 


Empirical studies on optimal interappointment times are commonly pursued in settings where analytical approaches face significant challenges. Accordingly, while we have proven optimality of the \hyperref[algcap]{ASAP} scheduling rule for the case of non-increasing costs, we derived further insights for the case of increasing costs by \textcolor{black}{analyzing optimal interappointment times in the Online Appendix~\ref{intapptimes}}. For non-increasing idle time costs, optimal interappointment times follow variants of the well-known plateau-dome pattern \citep{denton2003sequential, robinson2003scheduling, kong2020appointment, sang2021appointment}, with an initial increase, a plateau, and a decline toward the end of the horizon. In contrast, the structural complexity observed under increasing costs supports the hypothesis from Remark~\ref{remark_inc} that deriving optimal scheduling rules in this setting is more complex.

%
\section{Computational Study}
\label{subsection-experiments}
In this section, we evaluate the performance of \hyperref[RASWTG-k]{RASWTG} schedules on a real-world data set from the radiology department of a large university hospital. In particular, we analyze runtimes, conduct a sensitivity analysis, and present managerial insights from comparing waiting times, idle times, and overtime obtained with \hyperref[RASWTG-k]{RASWTG} to those obtained with a weighted sum approach. We use the standard time notation mm:ss, where mm denotes the number of completed minutes and ss the number of completed seconds since the start of the minute.

\subsection{Instances}

We generate test instances based on 7,745 CT examination records performed on one of three CT machines in the radiology department of a large university hospital in Munich, Germany. All exams were performed in 2019 or the first quarter of 2020. For each CT examination, we have the following information: Exam type, exam day, start time, and completion time. The start and completion times were recorded as so-called ``Modality Performed Procedure Steps''. That is, they do not include setup times before and after the irradiation procedure, such as preparing the patient at the unit or removing the patient from the unit, but only the scanning time. Service times are calculated as the difference between completion time and start time. 

We first split the data into an interval estimation set and a test set. The interval estimation set consists of all 6,359 CT examinations recorded in 2019; the test set consists of all 1,386 CT examinations recorded in the first quarter of 2020.
The interval estimation set is used to estimate the service time intervals for the examinations in the test set. Specifically, for each of the 64 examination types, we estimate the minimum and maximum service time as the 5th and 90th percentiles of the empirical distribution of service times within the interval estimation set. We tested that this choice ensures meeting the waiting time guarantees for nearly all patients while also resulting in reasonable total costs. We then compute schedules for the examination requests recorded in the test set using these service time interval estimates. In case of multiple optimal solutions, we choose one that minimizes the sum of appointment times in accordance with the \hyperref[algcap]{ASAP} rule. The schedules are evaluated using the service times recorded in the test set and a random sample of $n-k$ no-shows.

To generate test instances, we consider four factors: number of patients, level of no-shows, waiting time guarantees, and structure of the idle time costs. In a ceteris paribus design, we systematically vary these factors using three values each, as outlined in Table~\ref{tab_parameters}. Further insights into the rationale behind selecting these factors are available in the following and the \hyperref[AppLit]{Online Appendix}~\ref{instanceGen}. Unless specified otherwise, our analysis centers on the base case characterized by bold values, encompassing constant idle time costs, ten patients who all show up, and a maximum waiting time guarantee of 30 minutes.

\begin{table}[h]
\caption{Parameter Sets \normalfont{(Base Case in Bold)}}
\centering \footnotesize
\begin{tabular}{cccccccc}
\hline
{Number of patients} && {No-show rate} && {Waiting-time guarantee} && {Idle time cost}
\\ \cmidrule{1-1}\cmidrule{3-3}\cmidrule{5-5}\cmidrule{7-7}
$\lbrace 5, \,\textbf{10},\, 20 \rbrace $ && $\left\lbrace \textbf{0\%}, \, 10\%,\, 20\% \right\rbrace$ && $ \lbrace \textrm{10 min},\, \textrm{20 min}, \, \textrm{\bf 30 min} \rbrace $ && \{\textbf{const.}, dec., inc.\}\\ \hline
\end{tabular}
\label{tab_parameters}
\end{table}

For each $n \in \lbrace 5, 10, 20 \rbrace$, patient selection is performed by first identifying days within the test set that contain at least $n$ appointments. Subsequently, we randomly sample $n$ consecutive patient records from these days. This process generates 252 instances across all classes with five, 107 instances across all classes with ten, and 39 instances across all classes with twenty patients. 

We differentiate three types of idle time costs: constant, decreasing, and increasing. Constant idle time costs $(c_i)_{i \in[n]}$ remain fixed at $c_i = 1$ for all $i \in [n+1]$. Decreasing idle time costs follow three assumptions: Idle time before the first appointment incurs a per-unit time cost of $1$, idle time after the last appointment, fully utilized by staff for tasks like documentation, incurs a per-unit time cost of only $0.5$, and costs linearly decrease between the first and last idle time. This results in $c_i = 1 - \frac{i-1}{2n}$ for idle time before appointment $i \in [n+1]$. For completeness, we also consider linearly increasing idle time costs with $c_i = \frac{n+i-1}{2n}$, $i \in [n+1]$. Overtime cost, complying with current US federal law requiring a minimum of 1.5 times the regular pay rate, is set at $c_o = 1.25$.

Finally, the length $L$ of the planning horizon crucially impacts the trade-off between idle time cost and overtime cost. A short planning horizon may cause significant worst-case overtime, while an extended one may lead to no overtime but substantial idle time. For a comprehensive evaluation of the model, we set the length of the planning horizon for each instance to a specific time between the earliest and latest possible completion times of the last appointment, accommodating both idle time and overtime. Specifically, $L$ is chosen as the sum of maximum service times minus the waiting time guarantee for the last appointment.
For instances with zero no-shows and a waiting time guarantee of 30 minutes, the resulting average planning horizon is 50:05 minutes for five patients, 132:08 minutes for ten patients, and 312:32 minutes for twenty patients. The average cumulative uncertainty within this horizon, obtained as the sum of the service time interval lengths, amounts to 46:53 minutes for five patients, 91:02 minutes for ten patients, and 178:09 minutes for twenty patients, and accordingly higher in the case of no-shows.

\subsection{Runtimes}

In Table~\ref{tab_run}, we report the average runtimes to compute an optimal schedule using the polynomial-time algorithm (PTA) that results from first applying Lemma~\ref{opt-sequence-general} and then Algorithm~\ref{algcap} in the case of constant idle time costs and zero no-shows, the mixed-integer program \hyperref[RASWTG-n]{RASWTG-$0$} for the case of zero no-shows, and the general mixed-integer program \hyperref[RASWTG-k]{RASWTG}. All experiments have been performed on an Intel Core i7-10510U 1.8 GHz machine (in 64-bit mode) with 16 GB memory under Microsoft Windows 10 and Gurobi 9.5.0 in Python 3.9.0.

\begin{table}[h]
\caption{Average Runtimes to Compute an Optimal Schedule \normalfont{(in Seconds)}}
\centering \footnotesize
\begin{tabular}{lrrrcccccrrrrrrrrrr}
\hline
&&&& \multicolumn{3}{c}{Method} && \multicolumn{5}{c}{Patients} \\ \cmidrule{5-7} \cmidrule{9-13}
\multicolumn{1}{l}{Costs} && \multicolumn{1}{c}{No-shows} & & \hyperref[RASWTG-k]{RASWTG} & \hyperref[RASWTG-n]{RASWTG-$0$} & PTA && \multicolumn{1}{c}{5} && \multicolumn{1}{c}{10} && \multicolumn{1}{c}{20} \\ \cmidrule{1-7}\cmidrule{3-3}\cmidrule{5-5}\cmidrule{7-7}\cmidrule{9-9}\cmidrule{11-11}\cmidrule{13-13}
Constant && 0\% & & && $ \bullet $	&& 0.00 && 0.00 && 0.00 \\
Constant && 10\% & & $ \bullet $	&&&& 0.28 && 123.41 && - \\
Constant && 20\% & & $ \bullet $	&& && 0.28 && 92.80 && - \\
Decreasing & & 0\% &&& $ \bullet $ & && 0.01	&& 0.45 && 197.35 \\
Decreasing &&10\% && $ \bullet $ && && 0.45 && 799.21 && - \\
Decreasing &&20\% && $ \bullet $ && && 0.45 && 588.82 && -	\\
Increasing &&0\%& & & $ \bullet $ & && 0.05	&& 2.80 && 199.06 \\
Increasing &&10\%&& $ \bullet $ && && 0.72	&& 1993.71 && - \\
Increasing &&20\% && $ \bullet $ && && 0.72	&& 2541.86 && -
\\ \hline
\end{tabular}
\label{tab_run}
\end{table}

In the case of zero no-shows and constant idle time costs, our polynomial-time algorithm efficiently computes optimal schedules in milliseconds, even for a substantial number of patients. When there are zero no-shows but non-constant idle time costs, we employ the mixed-integer linear program \hyperref[RASWTG-n]{RASWTG-$0$}. With this model, Gurobi finds optimal schedules within seconds for up to 10 patients and within 200 seconds for 20 patients.

For the case of a positive number of no-shows, we utilize the mixed-integer linear program \hyperref[RASWTG-k]{RASWTG} to calculate optimal schedules. Computation times increase noticeably with higher patient numbers and no-show rates. The model remains computationally tractable for 10 patients. For situations involving 20 patients and a positive number of no-shows, Gurobi solves (almost) no instance within a one-hour time limit. This underscores the need for future research to develop scalable algorithms for addressing the robust appointment scheduling problem with no-shows. Furthermore, our experimental results support the conjecture from Section~\ref{subsection-rules} that decreasing idle time costs lead to simpler optimal solution structures than increasing idle time costs.

\subsection{Results}

We first conduct a sensitivity analysis and evaluate idle time and overtime costs relative to those obtained using a sample average approximation. Furthermore, we compare the RASWTG results with those achieved by minimizing a weighted sum of waiting times, idle times, and overtime.

\subsubsection{Sensitivity Analysis.} \label{sensana}
We assess the impact of each of the four factors listed in Table~\ref{tab_parameters}, as well as a sample average approximation, on waiting times, idle times, and overtime from which we derive practical managerial recommendations. To comprehensively understand the dynamics,
we recall the trade-offs between waiting time and idle time until the last appointment, and the trade-off between overtime and idle time following the last appointment.

\paragraph{Base Case.}
In the base case, the average waiting time per patient is 10:55 minutes, and the waiting time guarantees are met for 97\% of the patients. The slight discrepancy from 100\% is due to the estimation of minimum and maximum service times derived from empirical service time data in the separate interval estimation set, excluding outliers. In cases where realized service times are outside the estimated range, the waiting time guarantees of a few later patients are unmet. Furthermore, the average total idle time amounts to 37:40 minutes and overtime to 1:46 minutes, relative to an average cumulative uncertainty of 91:02 minutes. Figure~\ref{waitpattern} illustrates the distribution of waiting and idle times by appointment.

\begin{table}[b]
\caption{Number of Patients}
\centering \footnotesize
\begin{tabular}{lccccccccccccc}
\hline
&& 5 patients && 10 patients && 20 patients 
\\ \cmidrule{3-3} \cmidrule{5-5} \cmidrule{7-7}
{Share of patients whose waiting time guarantee is met:} && 96.11\% && 96.88\% && 98.23\%
\\
{Average idle time before an appointment (mm:ss):} && 1:54 && 3:46 && 4:55
\\
{Average overtime (mm:ss):} && 7:55 && 1:46 && 0:13 \\ \hline
\end{tabular}
\label{tab:horizon}
\end{table}

\paragraph{Number of Patients.}
Table \ref{tab:horizon} summarizes our findings regarding waiting times, idle times, and overtime across a varying number of patients.
Generally, as the number of patients in the planning horizon increases, the \hyperref[RASWTG-k]{RASWTG} approach leads to shorter waiting times and longer idle times (see \hyperref[AppLit]{Online Appendix}, Section \ref{remark_no}). Moreover, given our choice of the planning horizon~$L$, overtime decreases as the number of patients rises.

\begin{table}[t]
\caption{Total Idle Time (Idle) in Minutes, Average Waiting Time ($\overline{\text{Wait}}$) in Minutes, \qquad\qquad\qquad\qquad\qquad Percentage of Patients for Whom the Waiting Time Guarantee is Met (WTG), Overtime (Over)}
\centering \footnotesize
\resizebox{\textwidth}{!}{
\begin{tabular}{lrrrrrrrrrrrrrrrrrrrrrrrrrr}
\hline
&& \multicolumn{7}{c}{{10-minute guarantee}} && \multicolumn{7}{c}{{20-minute guarantee}} && \multicolumn{7}{c}{{30-minute guarantee}} \\ \cmidrule{3-9} \cmidrule{11-17} \cmidrule{19-25}
{No-shows} && Idle && $\overline{\text{Wait}}$ && WTG && Over && Idle && $\overline{\text{Wait}}$ && WTG && Over && Idle && $\overline{\text{Wait}}$ && WTG && Over \\ \cmidrule{1-1} \cmidrule{3-3} \cmidrule{5-5} \cmidrule{7-7} \cmidrule{9-9} \cmidrule{11-11} \cmidrule{13-13} \cmidrule{15-15} \cmidrule{17-17} \cmidrule{19-19} \cmidrule{21-21} \cmidrule{23-23} \cmidrule{25-25}
${0\%}$ && 56:45 &&	2:44 &&	91.50\% && 0:51 &&	47:01 &&	6:06 &&	94.86\% && 1:07 &&	37:40 &&	10:55 &&	96.88\% && 1:46 \\
${10\%}$ && 55:50 &&	2:07 &&	94.39\% && 1:37 &&	47:10 &&	4:51 &&	96.92\% && 1:25 &&	37:31 &&	8:02 &&	98.04\% &&	1:52 \\
${20\%}$ && 57:34 &&	1:52 &&	94.95\% && 3:12 &&	47:06 &&	4:15 &&	96.17\% &&	2:06 &&	37:56 &&	6:46 &&	98.13\% &&	2:46
\\ \hline
\end{tabular}
}
\label{tab:noshow}
\end{table}

\paragraph{No-Shows.} Table~\ref{tab:noshow} demonstrates that higher no-show rates correspond to enhanced adherence to waiting time guarantees and slightly increased overtime. Total idle time is not significantly affected by higher no-show rates, which highlights the model's performance with no-shows. Looking at the increase from 0\% to 10\% and from 10\% to 20\%, waiting times on average decrease by 1:08 minutes, and the fulfillment of waiting time guarantees increases by 1.14\%. This pattern is driven by the elevated planning uncertainty associated with higher no-show rates, prompting the conservatism of \hyperref[RASWTG-k]{RASWTG}, which ensures that all waiting time guarantees are met even if exactly the patients included in the worst-case scenario for waiting times show up. Moreover, a 10\% increase in the no-show rate leads to a 0:42-minute rise in overtime, attributed to the planning horizon choice dependent on the number of show-ups. Finally, it is important to note that the \hyperref[RASWTG-k]{RASWTG} approach relies on precise predictions of the number of no-shows. If practitioners cannot accurately forecast this number, we recommend underestimating the total number of no-shows to ensure the fulfillment of all waiting time guarantees.

\begin{figure}[b]
\caption{Average Waiting and Idle Times by Appointment \normalfont{(in Minutes)}}
\label{waitpattern}\begin{subfigure}{8.4cm}
\caption{\EGT\rm By Waiting Time Guarantee}
\label{waitpattern3}
\begin{subfigure}{4cm}
\begin{tikzpicture}[scale = 1, font=\EGT\rm]
\draw[opacity=0.0] (0,0) -- (0,-1.4);
\begin{axis}[xlabel={\EGT\rm Appointment no.}, ylabel={\EGT\rm Waiting time},ymax=20,xtick={5,10}, ytick={0,5, 10, 15},legend style={at={(0.5,-0.6)},anchor=north},mark size=0.25pt,mark=o,no markers, axis lines = left,ylabel style={align=center},
every axis plot/.append style={thick},width=4cm,height=4cm]
\addplot[TUMdarkgray,every mark/.append style={fill=blue!80!black},thick] table[x index=0,y index=1,col sep=comma] {data/data25a.dat};
\addplot[TUMalblack, dotted,every mark/.append style={fill=blue!80!black},thick] table[x index=0,y index=2,col sep=comma] {data/data25a.dat};
\addplot[TUMalblack,every mark/.append style={fill=blue!80!black},thick] table[x index=0,y index=3,col sep=comma] {data/data25a.dat};
\end{axis}
\end{tikzpicture}
\end{subfigure}
\begin{subfigure}{4cm}
\begin{tikzpicture}[scale = 1, font=\EGT\rm]
\draw[opacity=0.0] (0,0) -- (0,-1.4);
\begin{axis}[xlabel={\EGT\rm Appointment no.}, ylabel={\EGT\rm Idle time},xmax = 12 ,ymax=20,xtick={5,10}, ytick={0,5, 10, 15},legend style={at={(0.5,-0.6)},anchor=north},mark size=0.25pt,mark=o,no markers, axis lines = left,ylabel style={align=center},
every axis plot/.append style={thick},width=4cm,height=4cm]
\addplot[TUMdarkgray,every mark/.append style={fill=blue!80!black},thick] table[x index=0,y index=1,col sep=comma] {data/data29a.dat};
\addplot[TUMalblack, dotted,every mark/.append style={fill=blue!80!black},thick] table[x index=0,y index=2,col sep=comma] {data/data29a.dat};
\addplot[TUMalblack,every mark/.append style={fill=blue!80!black},thick] table[x index=0,y index=3,col sep=comma] {data/data29a.dat};
\end{axis}
\end{tikzpicture}
\end{subfigure}

\centering
\begin{tikzpicture}
\node
at ($(0,0)$)
{
{\begin{tikzpicture}[scale = 1, font=\EGT\rm]
\begin{customlegend}[legend entries={\EGT\rm 10 minutes, 20 minutes, 30 minutes}]
\addlegendimage{black,draw=TUMdarkgray, thick}
\addlegendimage{black,draw=TUMalblack, thick, dotted}
\addlegendimage{black,draw=TUMalblack, thick}
\end{customlegend}
\end{tikzpicture}}
};
\end{tikzpicture}
\end{subfigure}
\begin{subfigure}{8.4cm}
\caption{\EGT\rm By Idle Time Cost}
\label{waitpattern1}
\begin{subfigure}{4cm}
\begin{tikzpicture}[scale = 1, font=\EGT\rm]
\draw[opacity=0.0] (0,0) -- (0,-1.4);
\begin{axis}[xlabel={\EGT\rm Appointment no.}, ylabel={\EGT\rm Idle time},xmax = 12 ,ymax=20,xtick={5,10}, ytick={0,5, 10, 15},legend style={at={(0.5,-0.6)},anchor=north},mark size=0.25pt,mark=o,no markers, axis lines = left,ylabel style={align=center},
every axis plot/.append style={thick},width=4cm,height=4cm]
\addplot[TUMdarkgray,every mark/.append style={fill=blue!80!black},thick] table[x index=0,y index=3,col sep=comma] {data/data27.dat};
\addplot[TUMalblack, dotted,every mark/.append style={fill=blue!80!black},thick] table[x index=0,y index=1,col sep=comma] {data/data27.dat};
\addplot[TUMalblack,every mark/.append style={fill=blue!80!black},thick] table[x index=0,y index=2,col sep=comma] {data/data27.dat};
\end{axis}
\end{tikzpicture}
\end{subfigure}
\begin{subfigure}{4cm}
\begin{tikzpicture}[scale = 1, font=\EGT\rm]
\draw[opacity=0.0] (0,0) -- (0,-1.4);
\begin{axis}[xlabel={\EGT\rm Appointment no.}, ylabel={\EGT\rm Waiting time},xmax = 11 ,ymax=20,xtick={5,10}, ytick={0,5, 10, 15},legend style={at={(0.5,-0.6)},anchor=north},mark size=0.25pt,mark=o,no markers, axis lines = left,ylabel style={align=center},
every axis plot/.append style={thick},width=4cm,height=4cm]
\addplot[TUMdarkgray,every mark/.append style={fill=blue!80!black},thick] table[x index=0,y index=3,col sep=comma] {data/data26.dat};
\addplot[TUMalblack, dotted,every mark/.append style={fill=blue!80!black},thick] table[x index=0,y index=1,col sep=comma] {data/data26.dat};
\addplot[TUMalblack,every mark/.append style={fill=blue!80!black},thick] table[x index=0,y index=2,col sep=comma] {data/data26.dat};
\end{axis}
\end{tikzpicture}
\end{subfigure}

\centering
\begin{tikzpicture}
\node
at ($(0,0)$)
{
{\begin{tikzpicture}[scale = 1, font=\EGT\rm]
\begin{customlegend}[legend entries={\EGT\rm Decreasing cost, Constant cost, Increasing cost}]
\addlegendimage{black,draw=TUMalblack, thick}
\addlegendimage{black,draw=TUMalblack, thick, dotted}
\addlegendimage{black,draw=TUMdarkgray, thick}
\end{customlegend}
\end{tikzpicture}}
};
\end{tikzpicture}
\end{subfigure}
\end{figure}

\paragraph{Waiting Time Guarantees.}
Practitioners should tailor waiting time guarantees to align with their customers' waiting time tolerance levels. Figure~\ref{waitpattern3} illustrates how waiting and idle times vary across the planning horizon for different waiting time guarantee choices. Table~\ref{tab:noshow} shows the corresponding total idle time, average waiting times, and adherence to the waiting time guarantee. Increasing the waiting time guarantee by 10 minutes, on average, improves the fulfillment of the waiting time guarantee by 1.90\%, boosts capacity utilization by 18.44\%, and maintains stability in terms of overtime. This substantial capacity utilization increase results from greater flexibility in terms of the waiting time constraints and, correspondingly, a shorter overall planning horizon.

\paragraph{Idle Time Costs.} The idle time costs control whether idle times occur rather early or late in the planning horizon and, oppositely, whether waiting times occur rather early or late. Decreasing idle time costs, by definition, particularly avoid early idle times. Thus, idle times predominantly occur later in the planning horizon, as can be seen in Figure \ref{waitpattern1} for the base case. The opposite is true for increasing idle time costs. Constant idle time costs do not define a preference for when idle time occurs. Notably, both total idle time and overtime are highest in the case of increasing idle time costs and significantly lower and closely aligned for constant and decreasing idle time costs. This observation suggests that the simple greedy rule which utilizes capacities as early as possible proves effective when minimizing total idle time. Furthermore, constant and decreasing idle time costs lead to later and, thus, potentially more productively usable idle times.

\subsubsection{Comparison of RASWTG and Sample Average Approximation.} 
\label{choice_uncertaintyset}
While box uncertainty sets effectively limit excessive waiting times, idle time, and overtime with minimal distributional assumptions, alternative uncertainty models may yield lower expected idle and overtime costs. To estimate this effect in our problem instances, we applied a sample average approximation (SAA) to the base-case setting. \textcolor{black}{ For each patient~$j$, we drew $N \in \{10, 25\}$ service time samples $p_j^1, \ldots, p_j^N$ based on the examination type. The full model specification and detailed results can be found in Appendix~\hyperref[tab:costSAA]{D.3}.
Generally, as the number of samples increases, the average total idle time, overtime, and total cost decrease slightly, accompanied by a modest reduction in the proportion of patients for whom  the waiting-time guarantees are met. Runtimes increase substantially, which we attribute to the sample-dependent completion times.
Interestingly, the cost reductions achieved by the SAA are only marginal relative to the worst-case approach used in RASWTG. We attribute this to the hard waiting-time constraints, which inherently influence idle time and overtime in the objective function. These findings indicate that focusing on the scalability of the waiting-time constraints through refined uncertainty sets may be more beneficial than focusing solely on expected-cost optimization.
}

\subsubsection{Comparison to a Weighted Sum Approach.} \label{compw}
In this section, we compare waiting times, idle times, and overtime acquired using \hyperref[RASWTG-k]{RASWTG} in contrast to those obtained using a weighted sum approach.
Unless stated otherwise, we use the \hyperref[RASWTG-k]{RASWTG} approach in the base case setting. In particular, we assume that all patients show up.
We refer to the following problem as Weighted Sum Robust Appointment Scheduling Problem (WSRAS).
\begin{align} \label{TRAS}
\!\min_{\substack{\pi,A}}\, & U & \\
\text{s.t.}\, & U \geq \sum\limits_{i = 1}^{n + 1} c_i\left(A_i - {C}_{i^*,i-1}\right)^+ + \sum\limits_{i = 1}^{n} c_w\left({C}_{i^*,i-1} - A_i \right)^+ + c_o\left({C}_{i^*n} - L \right)^+ & \forall i^* \in [n+1] \label{conWS} \\
& {C}_{i^*,i} = \max\lbrace{A_i,{C}_{i^*,i-1}}\rbrace + \begin{cases}
\sum_{j=1}^n \pi_{ij}\ubar{p}_j, & i < i^* \\
\sum_{j=1}^n \pi_{ij}\bar{p}_j, & i \geq i^*
\end{cases} & \forall i^* \in [n+1] \label{compWS1} \\
& \sum_{j=1}^n \pi_{ij} = 1, \sum_{i=1}^n \pi_{ij} = 1 & \forall i,j \in [n] \label{assWS} \\
& A_1 = 0, A_{n+1} = L, {C}_{i^*0} = 0 & \forall i^* \in [n+1]
\end{align}
The Objective Function~(\ref{TRAS}) minimizes the upper bound on total cost, as established in Constraint~(\ref{conWS}). Specifically, Constraint~(\ref{conWS}) upper bounds the sum of waiting times, weighted by constant cost $c_w$, idle times, weighted by sequence-dependent cost $c_i$, and overtime, weighted by $c_o$; while ensuring robustness with respect to the same worst-case scenarios as \hyperref[RASWTG-k]{RASWTG} (see Theorem~\ref{wc}). Constraint~(\ref{compWS1}) defines the completion times, while Constraint~(\ref{assWS}) reflects the assignment constraint known from the Robust Appointment Scheduling Problem with Waiting Time Guarantees. To enhance our understanding of optimal \hyperref[TRAS]{WSRAS} schedules, we briefly explore optimal interappointment times, as illustrated in Figure~\ref{dome} in the \hyperref[AppLit]{Online Appendix}.

Notably, \cite{mittal2014robust} study a related robust appointment scheduling problem that minimizes a weighted sum of idle time and waiting time costs under a collective worst-case scenario. Our works show, independently of each other and with different objective functions, that it is sufficient to consider the same $n + 1$ critical service time scenarios (see Theorem~\ref{wc}). 

\begin{figure}[b]
\caption{WSRAS Waiting Times by Appointment \normalfont{(in Minutes)}}\label{waitt}
\begin{subfigure}{5cm}
\caption{\normalfont{$c_w = 0.00001$}}
\begin{tikzpicture}[font=\EGT\rm]
\begin{axis}
[boxplot/draw direction=y, height=5.0cm, width=5.0cm,draw=TUMalblack, xlabel = Appointment no., ylabel = Waiting time,
ytick={0,10,20,30,40,50,60,70,80,90}, xtick={1,2,3,4,5,6,7,8,9,10}, ymax = 100,
extra y ticks={30}, 
extra y tick style={
grid=major, 
ticklabel style={
fill=gray!20, 
text=black, 
},
},
]
\addplot+[TUMalblack,dash pattern = on 1pt off 0pt, mark=*, mark size = 0.5pt, mark options = {fill = white},
boxplot prepared={
lower whisker=0,
lower quartile=0,
median = 0,
upper quartile=0,
upper whisker=0
},
] coordinates {
};
\addplot+[TUMalblack,dash pattern = on 1pt off 0pt, mark=*, mark size = 0.5pt, mark options = {fill = white},
boxplot prepared={
lower whisker=0,
lower quartile=0.95,
median=2.88,
upper quartile=5.28,
upper whisker= 11.58
},
] coordinates {(0,11.9) (0, 13.05) (0, 19.21) (0, 21.45) (0, 23.06) (0, 28.78) (0, 30)};
\addplot+[TUMalblack,dash pattern = on 1pt off 0pt, mark=*, mark size = 0.5pt, mark options = {fill = white},
boxplot prepared={
lower whisker=0,
lower quartile=4.37,
median = 7.4,
upper quartile=11.75,
upper whisker=21.92
},
] coordinates {(0,23) (0, 24.83) (0, 26.97) (0, 31.22) (0, 33.03) (0, 34.44) (0, 35.63)};
\addplot+[TUMalblack,dash pattern = on 1pt off 0pt, mark=*, mark size = 0.5pt, mark options = {fill = white},
boxplot prepared={
lower whisker=0,
lower quartile=8.52,
median = 13.43,
upper quartile=18.21,
upper whisker=32.85
},
] coordinates {(0,38.07) (0, 44.63) (0, 50.32)};
\addplot+[TUMalblack,dash pattern = on 1pt off 0pt, mark=*, mark size = 0.5pt, mark options = {fill = white},
boxplot prepared={
lower whisker=0,
lower quartile=12.32,
median = 18.21,
upper quartile=24.78,
upper whisker=42.85
},
] coordinates {(0,48.65) (0, 50.52) (0, 57.73)};
\addplot+[TUMalblack,dash pattern = on 1pt off 0pt, mark=*, mark size = 0.5pt, mark options = {fill = white},
boxplot prepared={
lower whisker=0,
lower quartile=15.38,
median =22.13,
upper quartile=28.23,
upper whisker=46.57
},
] coordinates {(0,53.12) (0, 60.22) (0, 65.85)};
\addplot+[TUMalblack,dash pattern = on 1pt off 0pt, mark=*, mark size = 0.5pt, mark options = {fill = white},
boxplot prepared={
lower whisker=0,
lower quartile=14.05,
median = 25.81,
upper quartile=34.63,
upper whisker=64.7
},
] coordinates {(0,66.47) (0, 72.98)};
\addplot+[TUMalblack,dash pattern = on 1pt off 0pt, mark=*, mark size = 0.5pt, mark options = {fill = white},
boxplot prepared={
lower whisker=0,
lower quartile=17.3,
median = 28,
upper quartile=39.87,
upper whisker=68.62
},
] coordinates {(0,80.35)};
\addplot+[TUMalblack,dash pattern = on 1pt off 0pt, mark=*, mark size = 0.5pt, mark options = {fill = white},
boxplot prepared={
lower whisker=0,
lower quartile=14.63,
median = 30.31,
upper quartile=43.6,
upper whisker=84.13
},
] coordinates {(0,87.45)};
\addplot+[TUMalblack,dash pattern = on 1pt off 0pt, mark=*, mark size = 0.5pt, mark options = {fill = white},
boxplot prepared={
lower whisker=0,
lower quartile=4.88,
median = 29.9,
upper quartile=46.37,
upper whisker=89.55
},
] coordinates {};
\end{axis}
\end{tikzpicture}
\end{subfigure}
\begin{subfigure}{5cm}
\caption{\normalfont{$c_w = 0.1$}}
\begin{tikzpicture}[font=\EGT\rm]
\begin{axis}
[boxplot/draw direction=y, height=5.0cm, width=5.0cm,draw=TUMalblack, xlabel = Appointment no., ylabel = Waiting time,
ytick={0,10,20,30,40,50,60,70,80,90}, xtick={1,2,3,4,5,6,7,8,9,10}, ymax = 100,
extra y ticks={30}, 
extra y tick style={
grid=major, 
ticklabel style={
fill=gray!20, 
text=black, 
},
},
]
\addplot+[TUMalblack,dash pattern = on 1pt off 0pt, mark=*, mark size = 0.5pt, mark options = {fill = white},
boxplot prepared={
lower whisker=0,
lower quartile=0,
median = 0,
upper quartile=0,
upper whisker=0
},
] coordinates {
(0,25.46) (0,28.01) (0,28.45) (0,30.41) (0,30.50) (0,31.26) (0,35.28) (0,37.53) (0, 38.33) (0,43.15) (0,53.14)};
\addplot+[TUMalblack,dash pattern = on 1pt off 0pt, mark=*, mark size = 0.5pt, mark options = {fill = white},
boxplot prepared={
lower whisker=0,
lower quartile=0,
median = 2.08,
upper quartile=4.61,
upper whisker=11.42
},
] coordinates {(0,13.1) (0, 15.9) (0, 25.3)};
\addplot+[TUMalblack,dash pattern = on 1pt off 0pt, mark=*, mark size = 0.5pt, mark options = {fill = white},
boxplot prepared={
lower whisker=0,
lower quartile=0,
median = 2.72,
upper quartile=7.93,
upper whisker=17.43
},
] coordinates {(0,19.9) (0, 23) (0, 27.52) (0, 49.71)};
\addplot+[TUMalblack,dash pattern = on 1pt off 0pt, mark=*, mark size = 0.5pt, mark options = {fill = white},
boxplot prepared={
lower whisker=0,
lower quartile=0,
median = 5.85,
upper quartile=11.4,
upper whisker=23.68
},
] coordinates {(0,29.9) (0, 32.68) (0, 51.32)};
\addplot+[TUMalblack,dash pattern = on 1pt off 0pt, mark=*, mark size = 0.5pt, mark options = {fill = white},
boxplot prepared={
lower whisker=0,
lower quartile=0,
median = 5.21,
upper quartile=16.08,
upper whisker=32.22
},
] coordinates {(0,65.9)};
\addplot+[TUMalblack,dash pattern = on 1pt off 0pt, mark=*, mark size = 0.5pt, mark options = {fill = white},
boxplot prepared={
lower whisker=0,
lower quartile=0,
median = 7.95,
upper quartile=18.42,
upper whisker=40.33
},
] coordinates {(0,71.08)};
\addplot+[TUMalblack,dash pattern = on 1pt off 0pt, mark=*, mark size = 0.5pt, mark options = {fill = white},
boxplot prepared={
lower whisker=0,
lower quartile=0,
median = 9.25,
upper quartile=23.29,
upper whisker=47.75
},
] coordinates {(0,75.58)};
\addplot+[TUMalblack,dash pattern = on 1pt off 0pt, mark=*, mark size = 0.5pt, mark options = {fill = white},
boxplot prepared={
lower whisker=0,
lower quartile=0,
median = 10.73,
upper quartile=26.37,
upper whisker=56.18
},
] coordinates {(0,80.2)};
\addplot+[TUMalblack,dash pattern = on 1pt off 0pt, mark=*, mark size = 0.5pt, mark options = {fill = white},
boxplot prepared={
lower whisker=0,
lower quartile=0,
median = 7.47,
upper quartile=26,
upper whisker=60.73
},
] coordinates {(0,80.93)};
\addplot+[TUMalblack,dash pattern = on 1pt off 0pt, mark=*, mark size = 0.5pt, mark options = {fill = white},
boxplot prepared={
lower whisker=0,
lower quartile=0,
median = 0.52,
upper quartile=16.52,
upper whisker=36.15
},
] coordinates {(0,42.27) (0, 44.15) (0, 47.04) (0, 51.17) (0,53.85) (0,60.12) (0,82.95) (0,91.45)};
\end{axis}
\end{tikzpicture}
\end{subfigure}
\begin{subfigure}{5cm}
\caption{\normalfont{$c_w = 1$}}
\begin{tikzpicture}[font=\EGT\rm]
\begin{axis}
[boxplot/draw direction=y, height=5.0cm, width=5.0cm,draw=TUMalblack, xlabel = Appointment no., ylabel = Waiting time,
ytick={0,10,20,30,40,50,60,70,80,90}, xtick={1,2,3,4,5,6,7,8,9,10}, ymax = 100,
extra y ticks={30}, 
extra y tick style={
grid=major, 
ticklabel style={
fill=gray!20, 
text=black, 
},
},
]
\addplot+[TUMalblack,dash pattern = on 1pt off 0pt, mark=*, mark size = 0.5pt, mark options = {fill = white},
boxplot prepared={
lower whisker=0,
lower quartile=0,
median = 0,
upper quartile=0,
upper whisker=0
},
] coordinates {};
\addplot+[TUMalblack,dash pattern = on 1pt off 0pt, mark=*, mark size = 0.5pt, mark options = {fill = white},
boxplot prepared={
lower whisker=0,
lower quartile=0,
median = 0,
upper quartile=2.6,
upper whisker=6.47
},
] coordinates {(0,6.67) (0, 7) (0, 7.91) (0, 12.3)};
\addplot+[TUMalblack,dash pattern = on 1pt off 0pt, mark=*, mark size = 0.5pt, mark options = {fill = white},
boxplot prepared={
lower whisker=0,
lower quartile=0,
median = 0,
upper quartile=4.12,
upper whisker=9.93
},
] coordinates {(0,11.78) (0, 12.1) (0, 13.48) (0, 15.67) (0,19.41)};
\addplot+[TUMalblack,dash pattern = on 1pt off 0pt, mark=*, mark size = 0.5pt, mark options = {fill = white},
boxplot prepared={
lower whisker=0,
lower quartile=0,
median = 0,
upper quartile=2.55,
upper whisker=5.53
},
] coordinates {(0,6.59) (0, 7.02) (0, 7.71) (0,8.37) (0, 9.03) (0, 9.85) (0, 11.68) (0,14.19) (0,14.9) (0, 15.87) (0, 17.25) (0, 18.97) (0,23.38)};
\addplot+[TUMalblack,dash pattern = on 1pt off 0pt, mark=*, mark size = 0.5pt, mark options = {fill = white},
boxplot prepared={
lower whisker=0,
lower quartile=0,
median = 0,
upper quartile=1.54,
upper whisker=3.64
},
] coordinates {(0,4.13) (0, 4.77) (0, 5.27) (0,5.8) (0, 6.57) (0, 7.12) (0, 7.93) (0,8.5) (0,11.04) (0, 11.57) (0, 15.07)};
\addplot+[TUMalblack,dash pattern = on 1pt off 0pt, mark=*, mark size = 0.5pt, mark options = {fill = white},
boxplot prepared={
lower whisker=0,
lower quartile=0,
median = 0,
upper quartile=0.55,
upper whisker=1.25
},
] coordinates {(0,1.86) (0, 2.8) (0, 3.30) (0,3.56) (0, 3.98) (0, 4.40) (0, 4.78) (0,6.62) (0,7.43) (0, 7.77) (0, 8.29) (0,8.82) (0,10.25) (0, 11.05) (0, 11.92)};
\addplot+[TUMalblack,dash pattern = on 1pt off 0pt, mark=*, mark size = 0.5pt, mark options = {fill = white},
boxplot prepared={
lower whisker=0,
lower quartile=0,
median = 0,
upper quartile=1.25,
upper whisker=3.39
},
] coordinates {(0,3.65) (0, 4.37) (0, 4.82) (0,5.1) (0, 5.57) (0, 6.47) (0, 6.93) (0,7.22) (0,8.23) (0, 8.95) (0, 12.78) (0,13.1) (0,15.05)};
\addplot+[TUMalblack,dash pattern = on 1pt off 0pt, mark=*, mark size = 0.5pt, mark options = {fill = white},
boxplot prepared={
lower whisker=0,
lower quartile=0,
median = 0,
upper quartile=2.43,
upper whisker=6.02
},
] coordinates {(0,6.89) (0, 7.47) (0,9.01) (0,9.51) (0, 9.82) (0, 12.53) (0, 17.2) (0,18.33)};
\addplot+[TUMalblack,dash pattern = on 1pt off 0pt, mark=*, mark size = 0.5pt, mark options = {fill = white},
boxplot prepared={
lower whisker=0,
lower quartile=0,
median = 0,
upper quartile=3.65,
upper whisker=8.48
},
] coordinates {(0,9.14) (0, 10.68) (0,11.07) (0,14.18) (0, 16.12) (0, 25.33)};
\addplot+[TUMalblack,dash pattern = on 1pt off 0pt, mark=*, mark size = 0.5pt, mark options = {fill = white},
boxplot prepared={
lower whisker=0,
lower quartile=0,
median = 0,
upper quartile=2.16,
upper whisker=4.71
},
] coordinates {(0,5.57) (0, 5.83) (0, 6.97) (0, 8.44) (0,10) (0,10.58) (0,11.59) (0,12.24) (0,13.39) (0,18.23) (0,18.76) (0,19.64) (0,20.13) (0,23.63) (0,40.61)};
\end{axis}
\end{tikzpicture}
\end{subfigure}
\begin{subfigure}{1cm}
\centering
\begin{tikzpicture}[font=\EGT\rm]
\end{tikzpicture}
\end{subfigure}
\end{figure}
\begin{figure}[b]
\caption{RASWTG Waiting Times by Appointment \normalfont{(in Minutes)}} \label{waitRASWTG}
\centering
\begin{tikzpicture}[font=\EGT\rm]
\begin{axis}
[boxplot/draw direction=y, height=5.0cm, width=5.0cm,draw=TUMalblack, xlabel = Appointment no., ylabel = Waiting time,
ytick={0,10,20,30,40,50,60,70,80,90}, xtick={1,2,3,4,5,6,7,8,9,10}, ymax = 100,
extra y ticks={30}, 
extra y tick style={
grid=major, 
ticklabel style={
fill=gray!20, 
text=black, 
},
},
]
\addplot+[TUMalblack,dash pattern = on 1pt off 0pt, mark=*, mark size = 0.5pt, mark options = {fill = white},
boxplot prepared={
lower whisker=0,
lower quartile=0,
median = 0,
upper quartile=0,
upper whisker=0
},
] coordinates {};
\addplot+[TUMalblack,dash pattern = on 1pt off 0pt, mark=*, mark size = 0.5pt, mark options = {fill = white},
boxplot prepared={
lower whisker=2.38,
lower quartile=5.53,
median = 7.3,
upper quartile=9.38,
upper whisker=15.13
},
] coordinates {(0,18.92) (0,22.37) (0, 25.75)};
\addplot+[TUMalblack,dash pattern = on 1pt off 0pt, mark=*, mark size = 0.5pt, mark options = {fill = white},
boxplot prepared={
lower whisker=7.87,
lower quartile=12.63,
median = 14.83,
upper quartile=18.95,
upper whisker=28.18
},
] coordinates {(0,29.55) (0, 32.32) (0, 35.37)};
\addplot+[TUMalblack,dash pattern = on 1pt off 0pt, mark=*, mark size = 0.5pt, mark options = {fill = white},
boxplot prepared={
lower whisker=9.6,
lower quartile=17.32,
median = 20.45,
upper quartile=23.68,
upper whisker=32.07
},
] coordinates {(0,33.73) (0, 36.13) (0, 40.02) (0,46.18)};
\addplot+[TUMalblack,dash pattern = on 1pt off 0pt, mark=*, mark size = 0.5pt, mark options = {fill = white},
boxplot prepared={
lower whisker=4.02,
lower quartile=13.83,
median = 17.55,
upper quartile=23.03,
upper whisker=35.85
},
] coordinates {(0,40.67) (0, 47.81) };
\addplot+[TUMalblack,dash pattern = on 1pt off 0pt, mark=*, mark size = 0.5pt, mark options = {fill = white},
boxplot prepared={
lower whisker=0,
lower quartile=9.28,
median = 13.13,
upper quartile=20.68,
upper whisker=37.55
},
] coordinates {(0,38.42) (0, 41.18)};
\addplot+[TUMalblack,dash pattern = on 1pt off 0pt, mark=*, mark size = 0.5pt, mark options = {fill = white},
boxplot prepared={
lower whisker=0,
lower quartile=4.22,
median = 9.4,
upper quartile=16.25,
upper whisker=33.47
},
] coordinates {(0,38.73) (0, 36.73) (0, 45.56)};
\addplot+[TUMalblack,dash pattern = on 1pt off 0pt, mark=*, mark size = 0.5pt, mark options = {fill = white},
boxplot prepared={
lower whisker=0,
lower quartile=0,
median = 4.7,
upper quartile=13.15,
upper whisker=32.9
},
] coordinates {(0,32.91) (0, 43.11)};
\addplot+[TUMalblack,dash pattern = on 1pt off 0pt, mark=*, mark size = 0.5pt, mark options = {fill = white},
boxplot prepared={
lower whisker=0,
lower quartile=0,
median = 0,
upper quartile=9.9,
upper whisker=24
},
] coordinates {(0,28.11) (0, 31.95) (0,35.5) (0,43.57)};
\addplot+[TUMalblack,dash pattern = on 1pt off 0pt, mark=*, mark size = 0.5pt, mark options = {fill = white},
boxplot prepared={
lower whisker=0,
lower quartile=0,
median = 0,
upper quartile=1.33,
upper whisker=2.97
},
] coordinates {(0,5.85) (0, 3.5) (0, 7.72) (0, 10.17) (0,12.38) (0,14.05) (0,15.42) (0,17.43) (0,19.47) (0,28.83) (0,55.93) };
\end{axis}
\end{tikzpicture}
\end{figure}

\paragraph{Waiting Times.} We compare, first, how to control waiting times with both approaches and,
second, the distribution of waiting times across all appointments. Applying the weighted sum
approach, we need to carefully tune the waiting time costs to manage waiting times. In contrast, when applying RASWTG,
the waiting time guarantees constitute input parameters which (a priori) limit all worst-case waiting times, and indirectly also the
realized waiting times. It is therefore easier to control waiting times with RASWTG than with the
weighted sum approach.

In addition, Figures~\ref{waitt} and~\ref{waitRASWTG} present boxplots illustrating patient waiting times depending on their position in the appointment schedule. These waiting times correspond to the optimal interappointment intervals depicted in Figure~\ref{dome} in the \hyperref[AppLit]{Online Appendix}. Notably, for modest waiting time costs ($c_w$), waiting times with the weighted sum approach exhibit an increasing trend until the penultimate or last appointment, and excessive waiting times for some later appointments. High waiting time costs lead to very short waiting times for all appointments. In contrast, the \hyperref[RASWTG-k]{RASWTG} approach demonstrates an initial (limited) increase in waiting times followed by a gradual decrease. It is worth mentioning that this decline can be mitigated by introducing uncertainty budgets. Furthermore, for comparable overall costs, waiting times are distributed more equitably across appointments with \hyperref[RASWTG-k]{RASWTG}, and excessive waiting times are less likely.

\begin{table}[t]
\caption{Cost Incurred by Idle Time and Overtime}
\centering \footnotesize
\begin{tabular}{lcccccccccccccccc}
\hline
&& \hyperref[TRAS]{WSRAS} && \hyperref[TRAS]{WSRAS} && \textbf{RASWTG} && \hyperref[TRAS]{WSRAS} 
\\
&& $c_w = 0.00001$ && $c_w = 0.1$ && && $c_w = 1$ 
\\ \cmidrule{3-3} \cmidrule{5-5} \cmidrule{7-7} \cmidrule{9-9}
{Share of patients waiting at most 30 minutes:} && 79.91\% && 92.43\% && \textbf{96.88\%} && 99.91\%
\\
{Idle time up to the last appointment:} && 10:07 && 24:45 && \textbf{17:11} && 37:30
\\
{Total idle time:} && 37:55 && 39:02 && \textbf{37:40} && 42:16 \\
{Overtime:} && 2:01 && 3:08 && \textbf{1:46} && 6:22 \\ \hline
\end{tabular}
\label{tab:cost}
\end{table}

\paragraph{Costs.}
We finally evaluate the cost of replacing the weighted sum approach with \hyperref[RASWTG-k]{RASWTG}. We consider that a healthcare provider replaces \hyperref[TRAS]{WSRAS} schedules with fixed waiting time cost (i)~$c_w = 0.00001$, (ii) $c_w = 0.1$, and (iii) $c_w = 1$ over all days. Our findings are summarized in Table~\ref{tab:cost}.

Let us first examine overtime. For \hyperref[TRAS]{WSRAS}, we observe that overtime decreases with lower waiting time costs, as lower waiting costs allow for scheduling appointments earlier. Surprisingly, \hyperref[RASWTG-k]{RASWTG} exhibits the lowest overtime, on average only 1:46 minutes. This suggests that, even with hard waiting time constraints, \hyperref[RASWTG-k]{RASWTG} permits slightly earlier scheduling of the last appointment even compared to the weighted sum approach with low waiting time cost. Overall, both approaches maintain reasonable average overtime within a 1 to 7-minute range, and minimal with \hyperref[RASWTG-k]{RASWTG}.

Considering total idle time, we find that it naturally increases with rising waiting time costs for \hyperref[TRAS]{WSRAS}. \hyperref[RASWTG-k]{RASWTG} displays the lowest total idle time, a result consistent with our findings regarding overtime. It is worth noting that idle time for both approaches could be reduced by opting for a shorter planning horizon~$L$ but at the expense of increased overtime, which we aimed to prevent in the first place. However, as a substantial amount of idle time remains between the last appointment and the end of the planning horizon~$L$ for both approaches, we also compare the idle time up to the last appointment, which is less dependent on the choice of~$L$. Again, for all waiting time costs except for $c_w = 0.00001$, which comes with excessive waiting times, \hyperref[RASWTG-k]{RASWTG} yields lower idle time than \hyperref[TRAS]{WSRAS}.

We conclude with the following rule of thumb. In environments where most customers experience acceptable waiting times, introducing \hyperref[RASWTG-k]{RASWTG} offers both a priori waiting time guarantees and total cost reduction. The cost reduction typically comes with a limited increase in average waiting times. In environments where many customers experience excessive waiting times, introducing \hyperref[RASWTG-k]{RASWTG} effectively prevents excessive waiting times with a minimum increase in total worst-case cost. In all cases, \hyperref[RASWTG-k]{RASWTG} effectively mitigated excessive waiting times and overtime.

\section{Conclusion}

This work was motivated by the lack of intraday appointment scheduling and sequencing approaches to precisely control waiting times. We introduce maximum waiting time guarantees to respect customers' waiting time tolerance even in a worst-case scenario of uncertain service times and no-shows. Hence, the established guarantees prevent negative effects of waiting times on customer dissatisfaction, perceived service quality, and loyalty shown in previous empirical research. At the same time, we minimize worst-case costs incurred by idle time and overtime.

We show that the presented appointment scheduling problem with waiting time guarantees is $\mathcal{NP}$-hard and find optimal scheduling and sequencing rules in polynomial time for practically relevant special cases. Assuming constant idle time costs and zero no-shows, we show optimality of a sequencing rule that schedules customers in ascending order with the amount of uncertainty about their service times and their waiting tolerance. This rule provides a deeper understanding of the optimality of the well-known SVF rule and extends it to consider waiting time guarantees. Moreover, for the case of non-increasing idle time costs, we show optimality of the ASAP scheduling rule that sets each appointment time as early as possible while fulfilling the waiting time guarantees. The presented sequencing and scheduling rules are intuitive and easy to implement.

Our case study with real data from the radiology department of a large hospital shows that robust appointment scheduling with waiting time guarantees not only prevents excessive waiting times. Compared to weighted sum approaches, our approach also provides more flexibility to control the timing of idle and waiting times. Most importantly, we introduce the waiting time guarantees with a minimal increase in worst-case costs, which in many cases is even negative. Thus, compared to weighted sum approaches, waiting time guarantees can even reduce costs without increasing the waiting time of the longest waiting customer.

\textcolor{black}{A scalability limitation arises from the conservatism of box uncertainty sets, which assume all customers may simultaneously require extreme service times. To address this, more flexible uncertainty sets—such as polyhedral or general convex sets—are needed. With richer service time data, incorporating (potentially distributionally robust) chance constraints presents another promising direction. Such extensions further increase model complexity—especially in the already intricate cost function—and may pose challenges for deriving optimal scheduling and sequencing rules.}


Hence, this work points to several avenues for future research. First, extending the framework to more general uncertainty sets while preserving analytical tractability seems intriguing. Second, it would be interesting to integrate individual no-show predictions into our robust framework. Third, the underlying concept of minimizing costs under robust service level guarantees is applicable to any service in which customer disutility increases sharply upon surpassing a critical threshold. Identifying such services and studying their corresponding problem variants seems promising.

\ACKNOWLEDGMENT{%
Carolin Bauerhenne has been funded by the Deutsche Forschungsgemeinschaft (DFG, German Research Foundation) under grant 277991500/GRK 2201. We thank Bernhard Renger of the Department of Diagnostic and Interventional Radiology at the School of Medicine and Klinikum Rechts der Isar at the Technical University of Munich for providing the data.  
}

\addcontentsline{toc}{section}{References}
\bibliographystyle{template/informs2024} 

\bibliography{literature} 

@article{bendavid2018developing,
  title={Developing an optimal appointment scheduling for systems with rigid standby time under pre-determined quality of service},
  author={Bendavid, Illana and Marmor, Yariv N and Shnits, Boris},
  journal={Flexible Services and Manufacturing Journal},
  volume={30},
  number={1},
  pages={54--77},
  year={2018},
  publisher={Springer}
}

@misc{brugmanrobust,
  author       = {Brugman, Judith and den Hertog, Dick and van Leeuwaarden, Johan SH},
  title        = {Robust Appointment Scheduling with General Convex Uncertainty Sets},
  year         = {2024},
  note         = {Available at Optimization Online: \url{https://optimization-online.org/?p=27791}}
}

@article{kong2013scheduling,
  title={Scheduling arrivals to a stochastic service delivery system using copositive cones},
  author={Kong, Qingxia and Lee, Chung-Yee and Teo, Chung-Piaw and Zheng, Zhichao},
  journal={Operations Research},
  volume={61},
  number={3},
  pages={711--726},
  year={2013},
  publisher={INFORMS}
}

@article{millhiser2015designing,
  title={Designing appointment system templates with operational performance targets},
  author={Millhiser, William P and Veral, Emre A},
  journal={IIE Transactions on Healthcare Systems Engineering},
  volume={5},
  number={3},
  pages={125--146},
  year={2015},
  publisher={Taylor \& Francis}
}

@article{wang2014service,
  title={Service systems with finite and heterogeneous customer arrivals},
  author={Wang, Rowan and Jouini, Oualid and Benjaafar, Saif},
  journal={Manufacturing \& Service Operations Management},
  volume={16},
  number={3},
  pages={365--380},
  year={2014},
  publisher={INFORMS}
}

@article{millhiser2012assessing,
  title={Assessing appointment systems’ operational performance with policy targets},
  author={Millhiser, William P and Veral, Emre A and Valenti, Benedetto C},
  journal={IIE Transactions on Healthcare Systems Engineering},
  volume={2},
  number={4},
  pages={274--289},
  year={2012},
  publisher={Taylor \& Francis}
}

@article{jiang2019data,
  title={Data-driven distributionally robust appointment scheduling over Wasserstein balls},
  author={Jiang, Ruiwei and Ryu, Minseok and Xu, Guanglin},
  journal={arXiv preprint arXiv:1907.03219},
  year={2019}
}

@article{el1997robust,
  title={Robust solutions to least-squares problems with uncertain data},
  author={El Ghaoui, Laurent and Lebret, Herv{\'e}},
  journal={SIAM Journal on Matrix Analysis and Applications},
  volume={18},
  number={4},
  pages={1035--1064},
  year={1997},
  publisher={SIAM}
}

@article{el1998robust,
  title={Robust solutions to uncertain semidefinite programs},
  author={El Ghaoui, Laurent and Oustry, Francois and Lebret, Herv{\'e}},
  journal={SIAM Journal on Optimization},
  volume={9},
  number={1},
  pages={33--52},
  year={1998},
  publisher={SIAM}
}

@article{ben1998robust,
  title={Robust convex optimization},
  author={Ben-Tal, Aharon and Nemirovski, Arkadi},
  journal={Mathematics of Operations Research},
  volume={23},
  number={4},
  pages={769--805},
  year={1998},
  publisher={INFORMS}
}

@article{ben2000robust,
  title={Robust solutions of linear programming problems contaminated with uncertain data},
  author={Ben-Tal, Aharon and Nemirovski, Arkadi},
  journal={Mathematical Programming},
  volume={88},
  pages={411--424},
  year={2000},
  publisher={Springer}
}

@book{ben2009robust,
  title={Robust optimization},
  author={Ben-Tal, Aharon and El Ghaoui, Laurent and Nemirovski, Arkadi},
  volume={28},
  year={2009},
  publisher={Princeton university press}
}

@article{lu2021review,
  title={A review of robust operations management under model uncertainty},
  author={Lu, Mengshi and Shen, Zuo-Jun Max},
  journal={Production and Operations Management},
  volume={30},
  number={6},
  pages={1927--1943},
  year={2021},
  publisher={SAGE Publications Sage CA: Los Angeles, CA}
}

@article{jafarnia2020non,
  title={Non-indexability of the stochastic appointment scheduling problem},
  author={Jafarnia-Jahromi, Mehdi and Jain, Rahul},
  journal={Automatica},
  volume={118},
  pages={109016},
  year={2020},
  publisher={Elsevier}
}

@article{jouini2022appointment,
  title={Appointment-driven queueing systems with non-punctual customers},
  author={Jouini, Oualid and Benjaafar, Saif and Lu, Bingnan and Li, Siqiao and Legros, Benjamin},
  journal={Queueing Systems},
  volume={101},
  number={1},
  pages={1--56},
  year={2022},
  publisher={Springer}
}

@article{issabakhsh2021scheduling,
  title={Scheduling patient appointment in an infusion center: a mixed integer robust optimization approach},
  author={Issabakhsh, Mona and Lee, Seokgi and Kang, Hyojung},
  journal={Health Care Management Science},
  volume={24},
  number={1},
  pages={117--139},
  year={2021},
  publisher={Springer}
}

@article{gupta2008appointment,
  title={Appointment scheduling in health care: Challenges and opportunities},
  author={Gupta, Diwakar and Denton, Brian},
  journal={IIE Transactions},
  volume={40},
  number={9},
  pages={800--819},
  year={2008},
  publisher={Taylor \& Francis}
}

@article{deceuninck2018outpatient,
  title={Outpatient scheduling with unpunctual patients and no-shows},
  author={Deceuninck, Matthias and Fiems, Dieter and De Vuyst, Stijn},
  journal={European Journal of Operational Research},
  volume={265},
  number={1},
  pages={195--207},
  year={2018},
  publisher={Elsevier}
}

@book{garey1979computers,
  title={Computers and intractability: A guide to the theory of NP-hardness},
  author={Garey, Michael R and Johnson, David S},
  year={1979},
  publisher={San Fransisco: WH Freeman},
volume={1}
}

@article{bosch1997scheduling,
  title={Scheduling and sequencing arrivals to a stochastic service system.},
  author={Bosch, Peter Maurice Vanden},
  note={Doctoral Dissertation, Air Force Institute of Technology. ProQuest Dissertations and Theses Database (AAT 9813076)},
  year={1997},
volume={1}
}

@article{kong2016appointment,
  title={Appointment sequencing: Why the smallest-variance-first rule may not be optimal},
  author={Kong, Qingxia and Lee, Chung-Yee and Teo, Chung-Piaw and Zheng, Zhichao},
  journal={European Journal of Operational Research},
  volume={255},
  number={3},
  pages={809--821},
  year={2016},
  publisher={Elsevier}
}

@article{ahmadi2017outpatient,
  title={Outpatient appointment systems in healthcare: A review of optimization studies},
  author={Ahmadi-Javid, Amir and Jalali, Zahra and Klassen, Kenneth J},
  journal={European Journal of Operational Research},
  volume={258},
  number={1},
  pages={3--34},
  year={2017},
  publisher={Elsevier}
}

@article{cayirli2008assessment,
  title={Assessment of patient classification in appointment system design},
  author={Cayirli, Tugba and Veral, Emre and Rosen, Harry},
  journal={Production and Operations Management},
  volume={17},
  number={3},
  pages={338--353},
  year={2008},
  publisher={Wiley Online Library}
}

@article{taylor1994assessment,
  title={An assessment of the relationship between service quality and customer satisfaction in the formation of consumers' purchase intentions},
  author={Taylor, Steven A and Baker, Thomas L},
  journal={Journal of Retailing},
  volume={70},
  number={2},
  pages={163--178},
  year={1994},
  publisher={Elsevier}
}

@article{marynissen2019literature,
  title={Literature review on multi-appointment scheduling problems in hospitals},
  author={Marynissen, Joren and Demeulemeester, Erik},
  journal={European Journal of Operational Research},
  volume={272},
  number={2},
  pages={407--419},
  year={2019},
  publisher={Elsevier}
}

@article{bae2014assessing,
  title={Assessing the relationships between nurse work hours/overtime and nurse and patient outcomes: systematic literature review},
  author={Bae, Sung-Heui and Fabry, Donna},
  journal={Nursing outlook},
  volume={62},
  number={2},
  pages={138--156},
  year={2014},
  publisher={Elsevier}
}

@article{bertsimas2004price,
  title={The price of robustness},
  author={Bertsimas, Dimitris and Sim, Melvyn},
  journal={Operations Research},
  volume={52},
  number={1},
  pages={35--53},
  year={2004},
  publisher={Informs}
}

@article{bertsimas2011price,
  title={The price of fairness},
  author={Bertsimas, Dimitris and Farias, Vivek F and Trichakis, Nikolaos},
  journal={Operations Research},
  volume={59},
  number={1},
  pages={17--31},
  year={2011},
  publisher={INFORMS}
}

@article{zhang2017distributionally,
  title={Distributionally robust appointment scheduling with moment-based ambiguity set},
  author={Zhang, Yiling and Shen, Siqian and Erdogan, S Ayca},
  journal={Operations Research Letters},
  volume={45},
  number={2},
  pages={139--144},
  year={2017},
  publisher={Elsevier}
}

@article{jiang2017integer,
  title={Integer programming approaches for appointment scheduling with random no-shows and service durations},
  author={Jiang, Ruiwei and Shen, Siqian and Zhang, Yiling},
  journal={Operations Research},
  volume={65},
  number={6},
  pages={1638--1656},
  year={2017},
  publisher={INFORMS}
}

@article{qi2017mitigating,
  title={Mitigating delays and unfairness in appointment systems},
  author={Qi, Jin},
  journal={Management Science},
  volume={63},
  number={2},
  pages={566--583},
  year={2017},
  publisher={INFORMS}
}

@article{kong2020appointment,
  title={Appointment scheduling under time-dependent patient no-show behavior},
  author={Kong, Qingxia and Li, Shan and Liu, Nan and Teo, Chung-Piaw and Yan, Zhenzhen},
  journal={Management Science},
  volume={66},
  number={8},
  pages={3480--3500},
  year={2020},
  publisher={INFORMS}
}

@inproceedings{mittal2014robust,
  title={Robust appointment scheduling},
  author={Shashi Mittal and Andreas S. Schulz and Sebastian Stiller},
  booktitle={Approximation, Randomization, and Combinatorial Optimization. Algorithms and Techniques (APPROX/RANDOM 2014)},
  year={2014},
    pages =	{356--370},
  year =	{2014},
  volume =	{28},
  series =	{Leibniz International Proceedings in Informatics},
}

@article{de2021performance,
  title={Performance of the smallest-variance-first rule in appointment sequencing},
  author={de Kemp, Madelon A and Mandjes, Michel and Olver, Neil},
  journal={Operations Research},
  volume={69},
  number={6},
  pages={1909--1935},
  year={2021},
  publisher={INFORMS}
}

@inproceedings{schulz2019robust,
  author =	{Andreas S. Schulz and Rajan Udwani},
  title =	{{Robust appointment scheduling with heterogeneous costs}},
  booktitle={Approximation, Randomization, and Combinatorial Optimization. Algorithms and Techniques (APPROX/RANDOM 2014)},
    pages =	{25:1--25:17},
  number = {25},
  year =	{2019},
  volume =	{145},
  series =	{Leibniz International Proceedings in Informatics},
}

@article{zhou2019intraday,
  title={Intraday scheduling with patient re-entries and variability in behaviours},
  author={Zhou, Minglong and Loke, Gar Goei and Bandi, Chaithanya and Liau, Zi Qiang Glen and Wang, Wilson},
  journal={Manufacturing \& Service Operations Management},
  volume={24},
  number={1},
  pages={561--579},
  year={2022},
  publisher={INFORMS}
}

@article{benjaafar2020appointment,
  title={Appointment Scheduling Under a Service-Level Constraint},
  author={Benjaafar, Saif and Chen, David and Wang, Rowan and Yan, Zhenzhen},
  journal={Manufacturing \& Service Operations Management},
  volume={25},
  number={1},
  pages={70--87},
  year={2023},
  publisher={INFORMS}
}

@article{huang1994patient,
  title={Patient attitude towards waiting in an outpatient clinic and its applications},
  author={Huang, Xiao-Ming},
  journal={Health Services Management Research},
  volume={7},
  number={1},
  pages={2--8},
  year={1994},
  publisher={SAGE Publications Sage UK}
}

@article{samorani2020overbooked,
  title={Overbooked and overlooked: machine learning and racial bias in medical appointment scheduling},
  author={Samorani, Michele and Harris, Shannon L and Blount, Linda Goler and Lu, Haibing and Santoro, Michael A},
  journal={Manufacturing \& Service Operations Management},
  volume={24},
  number={6},
  pages={2825--2842},
  year={2022},
  publisher={INFORMS}
}

@article{cayirli2003outpatient,
  title={Outpatient scheduling in health care: a review of literature},
  author={Cayirli, Tugba and Veral, Emre},
  journal={Production and Operations Management},
  volume={12},
  number={4},
  pages={519--549},
  year={2003},
  publisher={Wiley Online Library}
}

@article{welch1952appointment,
  title={Appointment systems in hospital outpatient departments},
  author={Welch, JD and Bailey, NormanT J},
  journal={The Lancet},
  volume={259},
  number={6718},
  pages={1105--1108},
  year={1952},
  publisher={Elsevier}
}

@article{denton2003sequential,
  title={A sequential bounding approach for optimal appointment scheduling},
  author={Denton, Brian and Gupta, Diwakar},
  journal={IIE Transactions},
  volume={35},
  number={11},
  pages={1003--1016},
  year={2003},
  publisher={Taylor \& Francis}
}

@article{robinson2003scheduling,
  title={Scheduling doctors' appointments: optimal and empirically-based heuristic policies},
  author={Robinson, Lawrence W and Chen, Rachel R},
  journal={IIE Transactions},
  volume={35},
  number={3},
  pages={295--307},
  year={2003},
  publisher={Taylor \& Francis}
}

@article{davis1998disconfirmation,
  title={How disconfirmation, perception and actual waiting times impact customer satisfaction},
  author={Davis, Mark M and Heineke, Janelle},
  journal={International Journal of Service Industry Management},
   volume={9},
  number={1},
  pages={64--73},
  year={1998},
  publisher={MCB UP Ltd}
}

@article{bielen2007waiting,
  title={Waiting time influence on the satisfaction-loyalty relationship in services},
  author={Bielen, Fr{\'e}d{\'e}ric and Demoulin, Nathalie},
  journal={Managing Service Quality: An International Journal},
   volume={17},
  number={2},
  pages={174--193},
  year={2007},
  publisher={Emerald Group Publishing Limited}
}

@article{mak2015appointment,
  title={Appointment scheduling with limited distributional information},
  author={Mak, Ho-Yin and Rong, Ying and Zhang, Jiawei},
  journal={Management Science},
  volume={61},
  number={2},
  pages={316--334},
  year={2015},
  publisher={INFORMS}
}

@INBOOK{boydconvex,
  title={Convex Optimization},
  author={Boyd, Stephen and Vandenberghe, Lieven},
year={2004},
  publisher={Cambridge University Press},
pages={278},
volume={1}
}

@article{kumar1997impact,
  title={The impact of waiting time guarantees on customers' waiting experiences},
  author={Kumar, Piyush and Kalwani, Manohar U and Dada, Maqbool},
  journal={Marketing Science},
  volume={16},
  number={4},
  pages={295--314},
  year={1997},
  publisher={INFORMS}
}

@article{moschis2003influences,
  title={What influences the mature consumer?},
  author={Moschis, George P and Bellenger, Danny N and Curasi, Carolyn Folkman},
  journal={Marketing Health Services},
  volume={23},
  number={4},
  pages={16--21},
  year={2003}
}

@article{sang2021appointment,
  title={Appointment scheduling with a quantile objective},
  author={Sang, Peijun and Begen, Mehmet A and Cao, Jiguo},
  journal={Computers \& Operations Research},
  volume={132},
  pages={105295},
  year={2021},
  publisher={Elsevier}
}

@inproceedings{bouch2000quality,
  title={Quality is in the eye of the beholder: Meeting users' requirements for internet quality of service},
  author={Bouch, Anna and Kuchinsky, Allan and Bhatti, Nina},
  booktitle={Proceedings of the SIGCHI Conference on Human Factors in Computing Systems 2000},
  pages={297--304},
  year={2000},
volume={1}
}

@article{hill2005impact,
  title={The impact of unacceptable wait time on health care patients' attitudes and actions},
  author={Hill, C Jeanne and Joonas, Kishwar},
  journal={Health Marketing Quarterly},
  volume={23},
  number={2},
  pages={69--87},
  year={2005},
  publisher={Taylor \& Francis}
}

\newpage

 \begin{APPENDIX}{Online Supplement}
\noindent \emph{Article:} Robust Appointment Scheduling with Waiting Time Guarantees \\
 \emph{Authors:} Carolin Bauerhenne, Rainer Kolisch, Andreas S. Schulz  \\
 \emph{Appendix:} \label{AppLit}
The appendix is organized as follows: Section~\ref{A_proofs} presents additional proofs not included in the main paper. In Section~\ref{A_MILP}, we detail the step-by-step derivation of the general MILP from Section~\ref{ad-noshows}, considering no-shows. Section~\ref{intapptimes} provides further explanations of interappointment times from Section~\ref{interapp}. Lastly, Section~\ref{A_comp} covers details of the computational study, including instance generation, sensitivity analysis, and the comparison to the utilitarian approach.
 \vspace{10pt}
 
 \setlength{\extrarowheight}{.25em}
 \begin{table}[h]
 \centering \small
       \caption{\text{Notation}}
\begin{tabular}{llrll}  \toprule
    Index sets:    &&  $i \in [n]$ && appointments (sorted in the order of service receipt)  \\ 
         &&    $j \in [n]$ && customers     \vspace{0.125cm} \\ 
      Scenarios:   &&   $p \in \mathcal{P}$ & & service times  \\
             &&   $\lambda \in \Lambda_{k}$ & & no-shows   \vspace{0.125cm} \\ 
       Parameters:   &&  $L $& & length of the planning horizon \\ 
           &&   $k$& & number of customers showing up \\
    &&    $W_j$& & waiting time guarantee for customer $j$  \\
      &&  $c_i$ && unit cost for idle time before appointment $i$   \\
           &&   $c_o$ && unit cost for overtime   \vspace{0.125cm} \\ 
   Variables:   && $A_{i}$ && appointment time of appointment $i$   \\
     &&     $C_{i}$ && completion time of appointment $i$ \\
    && $\pi_{ij}$ && assignment of customer $j$ to appointment $i$  \\ \bottomrule
\end{tabular}
     \label{tab:notation}
 \end{table}

\section{Proofs} \label{A_proofs}
This section of the appendix contains the proofs for the results presented in the main part of this paper. In contrast to the main part, where the findings were organized for ease of understanding and applicability, here we structure the proofs to build upon previously established results.

First, we reformulate the definition of completion times, adopting a non-recursive computation approach. This reformulation allows for an essential transformation of idle times into non-negative blocks. Additionally, the reformulation of completion times facilitates to prove a crucial decomposition of worst-case scenarios for cost maximization, dividing them into two parts: the first one maximizing idle time and the other maximizing overtime. We demonstrate that considering $n+1$ service time scenarios is sufficient to maximize worst-case costs, and effectively linearize the search for corresponding worst-case no-show scenarios.
By leveraging these findings, we establish the ASAP scheduling rule, followed by the SVF-WTG sequencing rule, and ultimately prove $\mathcal{NP}$-hardness.

\begin{lemma} \label{completiontime}
For any service time scenario $p \in \mathcal{P}$ and no-show scenario $\lambda \in \Lambda_k$, the completion time of appointment $i \in [n]$ equals
\begin{align}
C_{i} = \max_{l \in [1,i]} \left ({A}_{l} + \sum\limits_{k = l}^i \sum\limits_{j = 1}^n \pi_{kj} \lambda_{j}p_{j} \right).
\end{align}
\end{lemma}

\proof{Proof of Lemma~\ref{completiontime}.} We prove by induction that the recursive computation of the completion times $\max\left\lbrace {A}_{i}, C_{i-1}\right\rbrace + \sum_{j = 1}^n \pi_{ij} \lambda_{j}p_{j}$ is equivalent to the maximization problem $\max_{l \in [1,i]} \left( {A}_{l} + \sum_{\sigma = l}^i \sum_{j = 1}^n \pi_{\sigma j} \lambda_{j}p_{j}\right)$. For $i = 1$, the terms are equivalent. Suppose the claim is true for all appointments $1, \ldots, i-1$. Then the recursive definition of the completion time for appointment $i$ is equivalent to
$ \max \left\lbrace {A}_{i}, \max_{l \in [1,i-1]} \left ({A}_{l} + \sum_{\sigma= l}^{i-1} \sum_{j = 1}^n \pi_{\sigma j} \lambda_{j}p_{j} \right) \right\rbrace + \sum_{j = 1}^n \pi_{ij} \lambda_{j}p_{j}. $
Taking the last summand into the maximization problem renders this equivalent to the above maximization problem. \rightline{\qed} 
\endproof \vspace{10pt}

\proof{Proof of Lemma~\ref{sep} (page~\pageref{sep}).}
Given a customer sequence $\pi$, appointment times $A$, and waiting time guarantees $W_1, \ldots, W_n$, it clearly holds
\begin{align*}
& \max\limits_{\boldsymbol{p} \in \mathcal{P}, \boldsymbol{\lambda} \in \Lambda_{k-1}} \left\lbrace \max\limits_{i \in [n]} \left\{ C_{i-1}(A, \pi, \boldsymbol{p}, \boldsymbol{\lambda})
- A_i - \sum_{j=1}^n \pi_{ij} W_j \right\} \right\rbrace 
\\
\leq & \max\limits_{i \in [n]} \left\lbrace \max\limits_{\boldsymbol{p^{(i)}} \in \mathcal{P}, \boldsymbol{\lambda^{(i)}} \in \Lambda_{k-1}} \left\{ C_{i-1}(A, \pi, \boldsymbol{p^{(i)}}, \boldsymbol{\lambda^{(i)}})
- A_i - \sum_{j=1}^n \pi_{ij} W_j \right\} \right\rbrace.
\end{align*}

Assume the inequality holds strictly (i.e., with "$<$"). Let $i^*, \boldsymbol{p^*}, \boldsymbol{\lambda^*}$ maximize the right-hand side. Then, for any $A \geq 0$, we obtain
\begin{align*}
& C_{i^*-1}(A, \pi, \boldsymbol{p^{*}}, \boldsymbol{\lambda^{*}}) - A_{i^*} - \sum_{j=1}^n \pi_{i^*j} W_j \\
\leq & \max_{i \in [n]} \left\lbrace C_{i-1}(A, \pi, \boldsymbol{p^*}, \boldsymbol{\lambda^*}) - A_{i} - \sum_{j=1}^n \pi_{ij} W_j \right\rbrace  \\
\leq & \max_{\boldsymbol{p} \in \mathcal{P}, \boldsymbol{\lambda} \in \Lambda_{k-1}} \left\lbrace \max_{i \in [n]} \left\lbrace C_{i-1}(A, \pi, \boldsymbol{p}, \boldsymbol{\lambda}) - A_{i} - \sum_{j=1}^n \pi_{ij} W_j \right\rbrace \right\rbrace \\
< & \max\limits_{i \in [n]} \left\{ \max\limits_{{\boldsymbol p^{(i)}} \in \mathcal{P}, \boldsymbol{\lambda^{(i)}} \in \Lambda_{k-1}} \left\lbrace C_{i-1}(A, \pi, \boldsymbol{p^{(i)}}, \boldsymbol{\lambda^{(i)}}) - A_i - \sum_{j=1}^n \pi_{ij} W_j 
\right\} \right\rbrace \\
& = C_{i^*-1}(A, \pi, \boldsymbol{p^{*}}, \boldsymbol{\lambda^{*}}) - A_{i^*} - \sum_{j=1}^n \pi_{i^*j} W_j,
\end{align*}
which is a contradiction.

\rightline{\qed} 
\endproof \vspace{10pt}

\textcolor{black}{
\begin{remark} \label{proof00o} A monotonic increase in the appointment times is not guaranteed: 
    Consider three customers with service times of $\ubar{p}_j, \bar{p}_j = 10$ for all $j\in[3]$. Suppose the customers assigned to the first two appointments have a waiting time guarantee of 0 time units, while the customer assigned to the last appointment has a waiting time guarantee of 30 time units. Under constant idle time costs and no-shows, Lemma~\ref{opt-sequence-general} and Theorem~\ref{opt-times} will demonstrate that there exists an   optimal solution with $\pi = id$ and appointment times $A_1 = 0$, $A_2 = 10$, $A_3 = 0$. 
Using the decomposability of the waiting time constraints in Lemma~\ref{sep}, we find that $A_i$ is only lower bounded by $\max_{p \in \mathcal{P}, \lambda \in \Lambda_{k-1}} C_{i-1}(\pi, A, p, \lambda) - \sum_{j=1}^n \pi_{ij}W_j$ and $C_{i-1}$ does not depend on $A_l$ for $l\geq i$. Hence, it holds that one may choose appointment times such that $A_{i-1} > A_i$ holds if and only if $A_{i-1} > \max_{p \in \mathcal{P}, \lambda \in \Lambda_{k-1}} C_{i-1}(\pi, A, p, \lambda) - \sum_{j=1}^n \pi_{ij}W_j$. The right-hand term is equivalent to $\max_{p \in \mathcal{P}, \lambda \in \Lambda_{k-1}}\left\lbrace\max\left\lbrace A_{i-1},  C_{i-2}(\pi, A, p, \lambda)\right\rbrace + \sum_{j = 1}^n \pi_{i-1,j}\lambda_j{ p_j}\right\rbrace - \sum_{j=1}^n \pi_{ij}W_j$.
\end{remark} \vspace{10pt}
}

\begin{lemma} \label{idletimeref} Let $\pi$ be an assignment, $(A_i)_{i \in [n]}$ appointment times, $p \in \mathcal{P}$ a service time scenario, and $\lambda \in \Lambda_k$ a no-show scenario. Moreover, let $ \lbrace i \in [n]: (A_i - C_{i-1}) > 0 \rbrace =\lbrace i_1, \ldots, i_r \rbrace =: I $ with $i_1 \leq \ldots \leq i_r$ denote the index set of appointments before which positive idle time occurs.
Then the idle time cost $\sum_{i=1}^{n+1} c_i (A_i - C_{i-1})^+ $ is equivalent to $$
\sum_{l = 1}^{r-1} (c_{i_l} - c_{i_{l+1}}) A_{i_l} + c_{i_r}A_{i_r} - c_{i_1}A_{1} - \sum_{l = 1}^r {c_{i_l}} \sum_{{\sigma}=i_{l-1}}^{i_l-1} {\lambda}_{{\sigma}} {p}_{{\sigma}}.$$
\end{lemma}

\proof{Proof of Lemma~\ref{idletimeref} (page~\pageref{idletimeref}).} Assume w.l.o.g. that $\pi = id$.
For each $i \in I$, let $l(i)$ denote the smallest index such that $C_{i-1} = A_{l(i)} + \sum\limits_{\sigma=l(i)}^{i-1} {p}_{\sigma} {\lambda}_{\sigma}$ (see Lemma~\ref{completiontime}).

We show that $l(i_j) = i_{j-1}$ for all $j \in [r]$, $j \geq 2$, via a proof by contradiction.
First assume that $l(i_{j+1}) < i_{j}$, and for the purpose of simpler notation, let w.l.o.g. $j = 2$.
On the one hand, Lemma~\ref{completiontime} implies that $C_{i_2-1} \geq A_{i_1} + \sum_{{\sigma}=i_1}^{i_2-1} {p}_{\sigma} {\lambda}_{\sigma}$. On the other hand, it holds that $C_{i_2-1} = A_{l(i_2)} + \sum_{{\sigma}=l(i)}^{i_1-1} {p}_{\sigma} {\lambda}_{\sigma} + \sum_{{\sigma}=i_1}^{i_2-1} {p}_{\sigma} {\lambda}_{\sigma} \leq C_{i_1-1} + \sum_{{\sigma}=i_1}^{i_2-1} {p}_{\sigma} {\lambda}_{\sigma} $. Together, it follows that $C_{i_1-1} \geq A_{i_1}$, a contradiction to $i_1 \in I$. Therefore, $l(i_2) \geq i_1$. \\
Now assume that $l(i_2) > i_1$. On the one hand, as $l(i_2)$ is the smallest index with the above property, it must hold that $C_{i_2-1} > A_{i_1} + \sum_{{\sigma}=i_1}^{i_2-1} {p}_{\sigma} {\lambda}_{\sigma}$. On the other hand, because now $l(i_2) \notin I$, it holds that $A_{i} \leq C_{i-1}$ for all $i \in [i_1 +1, l(i_2)]$ and therefore $C_{i_2-1} = A_{l(i_2)} + \sum_{{\sigma}=l(i_2)}^{i_2-1} {p}_{\sigma} {\lambda}_{\sigma} \leq C_{l(i_2)-1} + \sum_{{\sigma}=l(i_2)}^{i_2-1} {p}_{\sigma} {\lambda}_{\sigma} = \max\lbrace C_{l(i_2)-2}, A_{l(i_2)-1}\rbrace + \sum_{{\sigma}=l(i_2)-1}^{i_2-1} {p}_{\sigma} {\lambda}_{\sigma} \leq \ldots \leq A_{i_1} + \sum_{{\sigma}=i_1}^{i_2-1} {p}_{\sigma} {\lambda}_{\sigma}$. Together this yields the contradiction $A_{i_1} \geq C_{i_2-1} - \sum_{{\sigma}=i_1}^{i_2-1} {p}_{\sigma} {\lambda}_{\sigma} > A_{i_1}$, from which it finally follows that $l(i_2) = i_1$.

The above argument for $l(i_2) \leq i_1$ can be equivalently applied to show that $l(i_1) = 1$, so we define $i_0 := 1$. Altogether, this yields:
\begin{align}
& \sum_{i=1}^{n+1} c_i \left(A_i - C_{i-1} \right)^+ \label{reformulation1} \\
= & \sum_{l = 1}^r c_{i_l} \left(A_{i_l} - C_{{i_l}-1} \right) \\
= & \sum_{l = 1}^r c_{i_l} \left(A_{i_l} - A_{i_{l-1}} - \sum_{k=i_{l-1}}^{i_l-1} {p}_k {\lambda}_k \right ) \\
= & \sum_{l = 1}^{r-1} \left(c_{i_l} - c_{i_{l+1}}\right) A_{i_l} + c_{i_r}A_{i_r} - c_{i_1}A_{i_0} - \sum_{l = 1}^r {c_{i_l}} \sum_{{\sigma}=i_{l-1}}^{i_l-1} {\lambda}_{{\sigma}} {p}_{{\sigma}} \label{reformulation4}
\end{align}

\rightline{\qed} 
\endproof \vspace{10pt}

\proof{Proof of Theorem~\ref{wc} (page~\pageref{wc}).} We show that any service time scenario that maximizes total cost can be decomposed into two parts: the first one maximizing idle time and the second one maximizing overtime. Assume w.l.o.g. that $\pi = id$.

First, using the reformulation of the completion times from Lemma~\ref{completiontime}, we find:
\begin{align}
& \sum_{i = 1}^{n + 1} c_i \left(A_i - C_{i-1}(\pi, A, p, \lambda) \right)^+ + c_{o}\left(C_n(\pi, A, p, \lambda) - A_{n + 1} \right)^+ \label{obj}\\
= & \sum_{i = 1}^{n} c_i \left(A_i - C_{i-1}(\pi, A, p, \lambda) \right)^+ + \max\left\lbrace c_{n+1}\left(A_{n+1} - C_n(\pi, A, p, \lambda) \right), c_{o}(C_n(\pi, A, p, \lambda) - A_{n + 1})\right\rbrace \\
= & c_1A_1 + \sum_{i = 2}^{n} c_i \left(A_i - \max_{l \in [i-1]}\left\{A_l + \sum_{k = l}^{i-1} \lambda_k p_k \right\} \right)^+ \label{max00} \\
& + \max\left\{ c_{n+1}\left(A_{n+1} - \max_{l \in [n]}\left\{A_l + \sum_{k = l}^{n} \lambda_k p_k \right\}\right), c_{o}\left(\max_{l \in [n]}\left\{A_l + \sum_{k = l}^{n}\lambda_k p_k \right\} - A_{n + 1}\right)\right\} \label{max}
\end{align}

Now, let $\lambda \in \Lambda_k$ and let the service time scenario $p^{opt} \in \mathcal{P}$ maximize total cost $\sum_{i = 2}^{n + 1} c_i \left(A_i - C_{i-1}(\pi, A, p, \lambda)\right)^+ + c_{o}\left(C_n(\pi, A, p, \lambda) - A_{n + 1} \right)^+$. We distinguish the following two cases.


\emph{Case 1: $A_{n+1} - C_n(\pi, A, p^{opt}, \lambda) \geq 0$.}

In this case, overtime cost equals zero. Therefore, setting ${p}_i^{opt}$ to $\underline{p}_i$ for each $i \in [n]$ will lead to an equal or higher value of (\ref{max00})--(\ref{max}).

\emph{Case 2: $C_n(\pi, A, p^{opt}, \lambda) - A_{n+1} \geq 0$.}

In this case, there are non-negative overtime costs. Let $l^* \in [n]$ be chosen minimally such that \newline $C_n(\pi, A, p^{opt}, \lambda) = A_{l^*} + \sum_{k = l^*}^{n}\lambda_k p_k^{opt}$. In particular, idle time cannot occur after the $l^*$th appointment. \textcolor{black}{Suppose there is idle time $A_i - C_{i-1}(\pi, A, p^{opt}, \lambda) > 0$ for some $i \in [l^*, ..., n]$. By definition of the completion times, it follows that $C_n(\pi, A, p^{opt}, \lambda) \geq A_i + \sum_{k = i}^n \lambda_k p_k^{opt}$. The positive idle time implies $A_i + \sum_{k = i}^n \lambda_k p_k^{opt} > C_{i-1}(\pi, A, p^{opt}, \lambda) + \sum_{k = i}^n \lambda_k p_k^{opt}$. From the definition of completion times, the latter is bounded by $C_{i-1}(\pi, A, p^{opt}, \lambda) + \sum_{k = i}^n \lambda_k p_k^{opt} \geq A_{l^*} + \sum_{k = l^*}^n \lambda_k p_k^{opt}$. Combining these inequalities, we find that $C_n(\pi, A, p^{opt}, \lambda) 
 > A_{l^*} + \sum_{k = l^*}^n \lambda_k p_k^{opt}$, which contradicts the choice of $l^*$.} Hence, setting $p^{opt}_i$ to $\bar{p}_i$ for all $i \geq l^*$ and to $\underline{p}_i$ for all $i \leq l^* - 1$ will again lead to an equal or higher value of (\ref{max00})--(\ref{max}).
\rightline{\qed} 
\endproof \vspace{10pt}

\begin{remark}[Additional Mixed-Integer Linear Reformulation Techniques]
\label{MIPrefs}
For each $i^* \in \lbrace 0, \ldots, n+1 \rbrace$ and $ i \in [n]$, let $z_{i^*i} \in \lbrace 0,1 \rbrace$ be auxiliary binary variables and $M \geq \max_{i \in [n]} \max\lbrace A_i, C_{i^*,i-1} \rbrace$ an upper bound for the latest realized start time of any appointment, for example,~$M := A_1 + \sum_{j=1}^n \bar{p}_j$. Then $C_{i^*i} = \max\left\lbrace{A_i, C_{i^*i-1}}\right\rbrace + \sum_{j=1}^n \lambda_{i^*ij}\ubar{p}_j$ can be linearized as follows:

\begin{align}
& C_{i^*i} \geq A_i + \sum_{j=1}^n \lambda_{ij}^{(i^*,1)}\ubar{p}_{j} & \forall i \in [n] \\
& C_{i^*i}\geq C_{i^*,i-1} + \sum_{j=1}^n \lambda_{ij}^{(i^*,1)}\ubar{p}_{j} & \forall i \in [n] \\
& C_{i^*i} \leq A_i + \sum_{j=1}^n \lambda_{ij}^{(i^*,1)}\ubar{p}_{j} + z_{i^*i}M & \forall i \in [n] \\
& C_{i^*i} \leq C_{i^*,i-1} + \sum_{j=1}^n \lambda_{ij}^{(i^*,1)}\ubar{p}_{j} + (1-z_{i^*i})M & \forall i \in [n]
\end{align}

\end{remark} \vspace{10pt}

\label{theoremproof}
\proof{Proof of Theorem~\ref{opt-times} (page~\pageref{opt-times}).} It is straightforward to see that Algorithm~\ref{algcap} runs in polynomial time with the fact that line 5 in Algorithm~\ref{algcap} constitutes a linear program with only $O(n)$ variables and constraints.

The waiting time constraints enforce that $A_{i} \geq \left( \max_{l \in [i-1]} A_{l} + U_l- \sum_{j=1}^n \pi_{ij} W_{j} \right)^+$ for all $i \in [2,n]$. We now show that, if $A_1 = 0$, then $A_{i} = \left( \max_{l \in [i-1]} A_{l} + U_l- \sum_{j=1}^n \pi_{ij} W_{j} \right)^+$ for $i \in [2,n]$ is an optimal solution in the case of non-increasing idle time costs.  

Let $A^*$ be optimal appointment times with $A^*_{i'} > \left( \max_{l \in [i'-1]} A_{l}^* + U_l- \sum_{j=1}^n \pi_{i'j} W_{j} \right)^+$ for some $i' \in [2,n]$. 
As $c_{i-1} - c_{i} \geq 0$ for all $i \in [2,n]$, and using the cost reformulation \ref{reformulation1}--\ref{reformulation4} in the proof of Lemma~\ref{idletimeref} with corresponding index set $\mathcal{I}$, we distinguish the following two cases. Firstly, if $i' \in \mathcal{I}$, then choosing $A^*_{i'} = \left( \max_{l \in [i'-1]} A_{l}^* + U_l- \sum_{j=1}^n \pi_{i'j} W_{j} \right)^+$ leads to a better objective value, contradicting the optimality of $A^*$. Secondly, if $i' \in \mathcal{I}$, then choosing $A^*_{i'} = \left( \max_{l \in [i'-1]} A_{l}^* + U_l- \sum_{j=1}^n \pi_{i'j} W_{j} \right)^+$ does not change the optimal solution value. Thus, it is optimal to schedule each appointment time $A_i$, $i \in [n]$, as early as possible.
\rightline{\qed} \endproof \vspace{10pt}

\begin{remark}[Increasing idle time costs] \label{remark_inc} The case of increasing idle time costs seems to be more difficult, as a balance must be found between the following strategies. On the one hand, the last appointment should be scheduled as early as possible to avoid both idle time and overtime. On the other hand, later idle times are more costly than early idle times and can only be prevented by scheduling later appointments with relatively long worst-case waiting times. To allow for longer worst-case waiting times for later appointments while still fulfilling all waiting time guarantees, early appointments may need to be scheduled later than necessary due to the waiting time constraints. This may delay the appointment time of the last appointment.
\end{remark} \vspace{10pt}

\label{lemmaproof}
\proof{Proof of Lemma~\ref{opt-sequence-general} (page~\pageref{opt-sequence-general}).} 
Since the idle time costs are constant and all customers show up, we know from a special case of Theorem~\ref{opt-times} that $A_1 = 0$ and $A_{i} = \left(\sum_{k=1}^{i-1} \sum_{j=1}^n \pi_{kj}^*\bar{p}_j - \sum_{j=1}^n \pi_{ij}^* W_{j} \right)^+$ are optimal appointment times for $i \in [2,n]$.

Consider an optimal sequence $\pi^*$ that does not adhere to the SVF-WTG sequencing rule. Then there exists $i', j', j'' \in [n]$ satisfying $\pi_{i',j'}^*, \pi_{i'+1,j''}^* = 1$ and $\bar{p}_{j'} - \ubar{p}_{j'} + W_{j'}(1+c_o) > \bar{p}_{j''} - \ubar{p}_{j''} + W_{j''}(1+c_o)$. To establish the optimality of the SVF-WTG sequencing rule, we prove that (iteratively) swapping the positions of customers $j'$ and $j''$ does not increase the optimal worst-case cost $ \max_{i^* \in \lbrace0, \ldots, n+1\rbrace} \sum_{i=1}^{i^*} c_i \left(A_i - \ubar{C}_{i-1}(\pi^*, A)\right)^+ + c_o \left( A_{i^*} + \sum_{i = i^*}^n \pi_{ij}^*\bar{p}_j - L \right)^+ =: opt $.

First, we consider the costs associated with the case $i^* \in \lbrace 0, n+1 \rbrace$. For $i^* = 0$, the corresponding cost amounts to $c_o \left( \sum_{i = 1}^n \pi_{ij}^*\bar{p}_j - L \right)^+$, and swapping two customers does not affect it. For $i^* = n + 1$, using the idle time cost reformulation \ref{reformulation4} from Lemma~\ref{idletimeref}, the corresponding cost is equivalent to $c_{n+1}(L - \sum_{i=1}^n \sum_{j=1}^n \pi_{ij}^*\ubar{p}_j)$, and swapping the position of two customers does not affect this value. 

For each $i^* \in [n]$, the associated cost is $ \sum_{i=1}^{i^*} (A_i - \ubar{C}_{i-1}(\pi^*, A))^+ + c_o \left( A_{i^*} + \sum_{k = i^*}^n \pi_{kj}^*\bar{p}_j - L \right)^+$. \textcolor{black}{If the overtime is negative, then the associated cost is less than or equal to that of the above case $i^* = n + 1$, and therefore does not need further consideration. If $A_{i^*} = 0$, the associated cost is less than or equal to that in the above case of $i^* = 0$, and is therefore not relevant for further consideration either.
Therefore, suppose that $A_{i^*} = \sum_{k=1}^{i^*-1} \sum_{j=1}^n \pi_{kj}^*\bar{p}_j - \sum_{j=1}^n \pi_{ij}^* W_{j} > 0$. Then the total cost becomes $\sum_{i=1}^{i^*} (A_i - \ubar{C}_{i-1}(\pi^*, A))^+ + c_o \left( \sum_{i=1}^{n} \sum_{j=1}^n \pi_{ij}^*\bar{p}_j - \sum_{j=1}^n \pi_{i^*j} W_{j} - L \right)$.} Using again the idle time cost reformulation \ref{reformulation4} from Lemma~\ref{idletimeref}, the sequence-dependent value is:
\begin{align*}
& \sum_{i=1}^{i^*} (A_i - \ubar{C}_{i-1}(\pi^*, A))^+ - c_o\sum_{j=1}^n \pi_{i^*j}^*W_j \\
= & A_{i^*} - \sum_{k=1}^{i^* -1}\sum_{j=1}^n \pi_{kj}^*\ubar{p}_j - c_o\sum_{j=1}^n \pi_{i^*j}^*W_j \\
= & \sum_{k=1}^{i^*-1} \sum_{j=1}^n \pi_{kj}^*\bar{p}_j - \sum_{j=1}^n \pi_{i^*j}^* W_{j} - \sum_{k=1}^{i^* -1}\sum_{j=1}^n \pi_{kj}^*\ubar{p}_j -c_o \sum_{j=1}^n \pi_{i^*j}^*W_j \\
= & \sum_{k=1}^{i^*-1} \sum_{j=1}^n \pi_{kj}^*(\bar{p}_j - \ubar{p}_j )- (1+c_o)\sum_{j=1}^n \pi_{i^*j}^*W_j.
\end{align*}


We now show that changing the sequence to $\pi_{i',j''}, \pi_{i'+1, j'} = 1$, the cost associated with each $i^* \in [n]$ does not exceed the optimal solution value obtained with $\pi^*$.

For each $ i^* > i' + 1$ or $ i ^* < i'$, the solution value does not change when changing the sequence to $\pi_{i',j''}, \pi_{i'+1, j'} = 1$. 
For $ i^* = i' + 1$, it is easy to see that switching the position of $j'$ and $j''$ improves the associated costs:
\begin{align*}
& \sum_{k=1}^{i^*-2} \sum_{j=1}^n \pi_{kj}^*(\bar{p}_j - \ubar{p}_j ) + \bar{p}_{j'} - \ubar{p}_{j'}- (1+c_o)W_{j''} \\
> & \sum_{k=1}^{i^*-2} \sum_{j=1}^n \pi_{kj}^*(\bar{p}_j - \ubar{p}_j ) + \bar{p}_{j''} - \ubar{p}_{j''}- (1+c_o)W_{j'} 
\end{align*}
We note that $i^* = i' + 1$ cannot have been the maximizer of the total worst-case cost, as this would contradict the optimality of $\pi^*$.

Finally, for $i^* = i'$, we find

\begin{align*}
& \sum_{k=1}^{i^*-1} \sum_{j=1}^n \pi_{kj}^*(\bar{p}_j - \ubar{p}_j )- (1+c_o)\sum_{j=1}^n \pi_{i^*j}^*W_j \\
= &\sum_{k=1}^{i^*-1} \sum_{j=1}^n \pi_{kj}^*(\bar{p}_j - \ubar{p}_j )- (1+c_o)W_{j'} \\
< & \sum_{k=1}^{i^*-1} \sum_{j=1}^n \pi_{kj}^*(\bar{p}_j - \ubar{p}_j ) + \bar{p}_{j'} - \ubar{p}_{j'} - ( \bar{p}_{j''} - \ubar{p}_{j''} + W_{j''}(1+c_o)) \\
\leq & \sum_{k=1}^{i^*} \sum_{j=1}^n \pi_{kj}^*(\bar{p}_j - \ubar{p}_j ) - W_{j''}(1+c_o) \\
\leq &\, opt - c_o \left( \sum_{j = 1}^n\bar{p}_j - L \right )
\end{align*}
and therefore, changing the sequence to $\pi_{i',j''}, \pi_{i'+1, j'} = 1$ does not increase the optimal solution value for the case of $i^* = i'$ either. 

Thus, (iteratively) applying the SVF-WTG rule does not increase the total worst-case cost.
\rightline{\qed} \endproof \vspace{10pt}

\label{NPproof}
\proof{Proof of Theorem~\ref{NP} (page~\pageref{NP}).} We establish the $\mathcal{NP}$-hardness of the \hyperref[RASWTG]{RASWTG} problem for the case that all customers show up, have identical waiting time guarantees $W$, and the idle time costs satisfy $c_1 = \ldots = c_{\sigma} > c_{\sigma +1} = \ldots = c_n$ for some $\sigma \in [n-1]$ and the overtime cost $c_o = 0$. To simplify the notation, let $\epsilon$ be a real number such that $L = \sum_{j=1}^n \bar{p}_j - W + \epsilon =: A_{n+1}$.

First, we employ Theorem~\ref{wc} to reformulate the objective function as
$\min_{\pi, A} \left\{ \max_{i^* \in \lbrace0, \ldots, n+1\rbrace} \sum_{i=1}^{i^*} c_i(A_i - \ubar{C}_{i-1}(\pi, A))^+ + c_o \left( A_{i^*} + \sum_{k = l^*}^n \sum_{j=1}^n \pi_{kj}\bar{p}_j - L \right)^+ \right\}$. Since the overtime cost fulfill $c_o = 0$, it suffices to consider $i^* \in \{ n+1 \}$. Moreover, using Theorem~\ref{opt-times}, the appointment start times can be calculated as $A_1 = 0$, $A_i = (\sum_{l=1}^{i-1}\sum_{j=1}^n \pi_{lj}\Bar{p}_j - W)^+$ for $i \in \lbrace 2, \ldots, n \rbrace$.
Therefore, the worst-case cost equals:
\begin{align*}
&\sum\limits_{i = 2}^{n+1} c_i\left(A_i-\ubar{C}_{i-1}(\pi, A) \right)^+ \\
=& \sum\limits_{i = 2}^{n+1} c_i\left(\sum\limits_{k=1}^{i-1}\sum\limits_{j=1}^n \pi_{kj}\bar{p}_j - W - \max\left\lbrace \sum_{k = 1}^{i-1} \sum_{j=1}^n \pi_{kj}\ubar{p}_j, \max_{l \in [i-1]} \left( \sum_{k = 1}^{l-1} \sum_{j=1}^n \pi_{kj}\bar{p}_j + \sum_{k = l}^{i-1} \sum_{j=1}^n \pi_{kj}{\ubar{p}_j} \right) -W \right\rbrace\right)^+ + c_{n+1}\epsilon & \\
= & \sum\limits_{i = 2}^{n+1} c_i \underbrace{\left(\min\left\lbrace \sum\limits_{k=1}^{i-1}\sum\limits_{j=1}^n \pi_{kj}(\bar{p}_j-\ubar{p}_j) - W,
\sum_{j=1}^n \pi_{i-1,j}(\bar{p}_{j}- {\ubar{p}_{j}})
\right\rbrace\right)^+}_{:= s_i} + c_{n+1}\epsilon \\
\end{align*}

We observe that $ \sum_{j=1}^n \pi_{i-1,j}(\bar{p}_{j}- {\ubar{p}_{j}}) \geq 0$ holds for all $i \in [2,{n+1}]$ in the second term in $s_i$ and that $s_i = 0$ if $\sum_{k=1}^{i-1}\sum_{j=1}^n \pi_{kj}(\bar{p}_j-\ubar{p}_j) - W \leq 0$. Let $l^*$ be the smallest index such that $\sum_{k=1}^{l^*-1}\sum_{j=1}^n \pi_{kj}(\bar{p}_j-\ubar{p}_j) - W \geq 0$. Then, for all $i > l^*$, it holds that
$
\sum_{j=1}^n \pi_{i-1,j}(\bar{p}_{j}- {\ubar{p}_{j}} )
\leq \sum_{k=1}^{l^*-1}\sum_{j=1}^n \pi_{kj}(\bar{p}_j-\ubar{p}_j) - W + \sum_{j=1}^n \pi_{i-1,j}(\bar{p}_{j}- {\ubar{p}_{j}} )
\leq \sum_{k=1}^{i-1}\sum_{j=1}^n \pi_{kj}(\bar{p}_j-\ubar{p}_j) - W.
$
Therefore, the worst-case cost finally becomes:
\begin{align}
& \sum\limits_{i = 1}^{n+1} c_i(A_i-\ubar{C}_{i-1})^+ \\
= & c_{l^*}\left(\sum\limits_{i=1}^{l^*-1}\sum\limits_{j=1}^n \pi_{ij}(\bar{p}_j-\ubar{p}_j) - W\right) + \sum\limits_{i = l^*+1}^{n+1}c_i \sum\limits_{j=1}^n \pi_{i-1,j}(\Bar{p}_{j}-\ubar{p}_{j}) + c_{n+1}\epsilon.
\end{align}
Now, we observe that the index $l^*$ is uniquely determined by the two conditions:
\begin{enumerate}
\item[(i)] $\sum\limits_{i=1}^{l^*-1}\sum\limits_{j=1}^n \pi_{ij}(\bar{p}_j-\ubar{p}_j) - W \geq 0$
\item[(ii)] $\sum\limits_{i=1}^{l^*-2}\sum\limits_{j=1}^n \pi_{ij}(\bar{p}_j-\ubar{p}_j) - W < 0$
\end{enumerate}
Notably, even if $\sum\limits_{i=1}^{l^*-2}\sum\limits_{j=1}^n \pi_{ij}(\bar{p}_j-\ubar{p}_j) - W = 0$, then the worst-case sum of idle time equals (60).
Therefore, the \hyperref[RASWTG]{RASWTG} problem under consideration becomes:
\begin{align}
\min_{l^* \in [2,{n+1}]} \min_{\pi} \quad& c_{l^*}\left(\sum\limits_{j=1}^n \sum\limits_{i = 1}^{l^*-1} \pi_{ij}(\bar{p}_j-\ubar{p}_j) \right) + \sum\limits_{j=1}^n \sum\limits_{i = l^*+1}^n \pi_{i-1,j}c_i(\Bar{p}_{j}-\ubar{p}_{j}) - c_{l^*}W +c_{n+1}\epsilon \label{np1} \\
\text{s.t.} \quad & W \leq \sum\limits_{i = 1}^{l^*-1}\sum\limits_{j=1}^n \pi_{ij} (\bar{p}_j-\ubar{p}_j) \leq W + \sum\limits_{j=1}^n \pi_{l^*-1,j}(\bar{p}_{j} -{\ubar{p}}_{j}) \label{np2} \\
& \sum\limits_{j = 1}^n \pi_{ij} = 1, \sum\limits_{i = 1}^n \pi_{ij} = 1\\
& \pi \in \lbrace 0,1\rbrace^{n \times n}   \label{np3}
\end{align}

We now reduce the following $\mathcal{NP}$-hard Subset Sum Problem \citep[p. 226]{garey1979computers} to a variant of (\ref{np1})~--~(\ref{np3}).
\begin{align}
\max_{z} \quad& \sum\limits_{j=1}^n \alpha_j z_j \\
\text{s.t.} \quad & \sum\limits_{j=1}^n \alpha_jz_j \leq \beta \\
& z \in \lbrace 0,1 \rbrace^n
\end{align}

Defining $y_j := 1 - z_j$ for all $j \in [n]$, the Subset Sum Problem is equivalent to the following problem. In particular, denoting the optimal value of the Subset Sum Problem by $OPT_z$ and the optimal value of the following problem by $OPT_y$, it holds $OPT_z = \sum_{j=1}^n \alpha_j - OPT_y$.
%
\begin{align}
\min_{l^* \in [0,n]} \min_{y} \quad& \sum\limits_{j=1}^n \alpha_j y_j \\
\text{s.t.} \quad & \sum\limits_{j=1}^n \alpha_j - \beta \leq \sum\limits_{j=1}^n \alpha_j y_j \\
& \sum\limits_{j=1}^n y_j = l^* \\
& y \in \lbrace 0,1 \rbrace^n
\end{align}

If the $l^*$th problem is infeasible, we assign it a huge number. Now, w.l.o.g. (the case $l^* = 0$ is trivial to check), the latter problem is equivalent to:

\begin{align}
\min_{l^* \in [n]} \min_{\pi} \quad& \sum\limits_{i = 1}^{l^*}\sum\limits_{j=1}^n \alpha_j \pi_{ij} \label{subs1} \\
\text{s.t.} \quad & \sum\limits_{j=1}^n \alpha_j - \beta \leq \sum\limits_{i = 1}^{l^*}\sum\limits_{j=1}^n \alpha_j \pi_{ij} \\
& \sum\limits_{j = 1}^n \pi_{ij} = 1, \sum\limits_{i = 1}^n \pi_{ij} = 1\\
& \pi_{ij} \in \lbrace 0,1 \rbrace^{n \times n} \label{subs2}
\end{align}

We can now use the similarity of (\ref{subs1}) -- (\ref{subs2}) and (\ref{np1}) -- (\ref{np3}).
Let $\alpha$ and $\beta$ be an instance for (\ref{subs1}) -- (\ref{subs2}). For $l^* \in [n]$ choose $\bar{p}$, $\ubar{p}$ fulfilling $\bar{p}_j - \ubar{p}_j := \alpha_j$ for each $j \in [n]$, $W := \sum\limits_{j=1}^n \alpha_j - \beta$, $c_1 = \ldots = c_{l^*} :=1$, and $c_{l^* +1} = \ldots = c_{n+ 1} := 0$. Then solving the $l^*$th subproblem of (\ref{subs1}) -- (\ref{subs2}) is equivalent to solving the respective $(l^*+1)$-st subproblem of (\ref{np1}) -- (\ref{np2}).
In particular, an optimal solution $(i_{OPT}, \pi_{OPT})$ also fulfills the second inequality in (\ref{np2}): If for some $j \in [n]$ and $l \leq i_{OPT}$ such that $\pi_{lj} = 1$ holds $\sum\limits_{i \in [1,i_{OPT}]\setminus \lbrace l \rbrace} \sum_{j= 1}^n \alpha_j \pi_{ij} > \sum\limits_{j=1}^n \alpha_j - \beta$, then there exists a solution $(l^*,\pi^*)$, $l^* < i_{OPT}$ with smaller solution value, which contradicts the optimality of $(i_{OPT}, \pi_{OPT})$.


Thus, the problem (\ref{np1}) -- (\ref{np3}), and therefore also the original problem, are $\mathcal{NP}$-hard.

\rightline{\qed}
\endproof \vspace{0pt}

\section{Mixed-Integer Linear Program} \label{A_MILP}
We now derive and present the mixed-integer linear program for the Robust Appointment Scheduling Problem with Waiting Time Guarantees, accounting for $0 \leq k \leq n$ no-shows. This involves linearizing the adversarial problems related to maximizing waiting time for each customer and the most challenging case of maximizing total cost.

\paragraph{Adversarial Problem of Maximizing Waiting Times.}
We first address the problem of describing a no-show scenario $ \argmax_{\lambda \in \Lambda_{k -1}} \left\lbrace \max_{l \in [1,i-1]} \left( {A}_{l} + \sum_{k = l}^{i-1} \sum_{j = 1}^n \pi_{kj} {\lambda}_{j}\bar{p}_{j} \right ) 
\right\rbrace $ that maximizes the completion time of the last appointment before appointment $i\in[n]$, to which the $j$-th customer is assigned (who shows up). Given the inner term $ {A}_{l} + \sum_{k = l}^{i-1} \sum_{j = 1}^n \pi_{kj} {\lambda}_{j}^{(i)}\bar{p}_{j} -A_i$, $l\in [i-1]$, $i \in [n]$, in a worst-case no-show scenario, both the $j$-th customer and as many as possible but at most $k-1$ of the previous $i-l$ customers show up, namely those with the longest maximum service times. Now, following \cite[p. 278]{boydconvex} for the equivalent max-$(k-1)$-sum problem, we can use total unimodularity of the constraint matrix as well as duality to see that the objective value of $ \max_{{{\lambda} \in \Lambda_{k-1}}} \sum_{k = l}^{i-1} \sum_{j = 1}^{n} \pi_{kj} {\lambda}_{j} \bar{p}_{j} $ equals $\min_{\alpha, z \geq 0} \left\lbrace (k-1)\,\alpha + \sum_{k=l}^{i-1} z_k \,|\, z_k + \alpha \geq \sum_{j = 1}^n \pi_{kj} \bar{p}_{j} \right\rbrace. $
Therefore, for all appointments $i \in [n]$, $l \in [i-1]$ such that $i-l \geq k$, the waiting time constraints in (\ref{rcon}) can be formulated as \begin{align}\label{waitnoshows}
A_l + \min_{\alpha, z \geq 0} \left\lbrace (k-1)\,\alpha + \sum_{j=1}^n z_j \,|\,  \alpha  + z_j\geq \sum\limits_{j = 1}^n \pi_{kj} \bar{p}_{j} \, \forall k\in [l,i-1] \right\rbrace - A_i \leq \sum_{j=1}^n\pi_{ij}W_j. \\ \notag
\end{align}

\paragraph{Adversarial Problem of Maximizing Cost.}
The problem of identifying the customers that show up when maximizing total cost is more difficult. This is because a no-show not only incurs costs proportional to the customer's service time but also involves sequence-dependent idle time and overtime costs. To tackle this issue, the following lemma characterizes a worst-case no-show scenario structure. 
Specifically, it becomes sufficient to consider, for each $i^* \in \lbrace 1, \ldots, n + 1 \rbrace$, a no-show scenario in which, among the customers scheduled for the last $n - i^* + 1$ appointments, those with the longest maximum service times show up. If $k > n - i^* + 1$, all customers scheduled for these last appointment show up. Additionally, among the customers scheduled for the first $i^* -1 $ appointments, the $k - (n - i^* + 1)$ ones with the smallest product of idle time cost and minimum service time show up. 

\begin{lemma} \label{wcc} Let $i^* \in \lbrace 1, \ldots, n+1 \rbrace$. For any no-show scenario ${\lambda} \in \Lambda_k$ and appointment $i \in [n]$, let $c(i) := c_{\min\lbrace{l \in[i,n]: A_l - \ubar{C}_{l-1}(\pi, A, \lambda) > 0 \rbrace}}$ denote the unit idle time cost of the next appointment before which a positive amount of idle time is incurred under minimum service times $\ubar{p}$. If no idle time is incurred for any of these appointments, then we choose $c(i) := \max_{l \in [n]}{c_l}$.

Then ${\lambda} \in \Lambda_k$ maximizes the total cost $\sum_{i=1}^{i^*} c_i(A_i - \ubar{C}_{i-1}(\pi, A, \lambda))^+ + c_o\left( A_{i^*} + \sum_{k = i^*}^n \pi_{kj}\lambda_j\bar{p}_j - L \right)^+$ if $\lambda$ fulfills the following conditions. In case that $k \leq n - i^* + 1$, then ${\lambda}_j = 1 $ for any $j \in [n]$ implies that $ \sum_{i=i^*}^n \pi_{ij} \bar{p}_j$ is among the $k$ largest components of $\left(\sum_{i=i^*}^{n} \pi_{ij} \bar{p}_j\right)_{j \in [n]}$. In case that $k > n - i^* + 1$, then it must hold $\sum_{j= 1}^n\pi_{ij}\lambda_j = 1$ for all $i \geq i^*$, and if $\pi_{ij}\lambda_j = 1$ for some $i < i^*$ and $j\in[n]$, then $ c(i) \ubar{p}_j$ is among the $k - (n - i^* + 1)$ smallest components of $\left(\sum_{i=1}^{i^* -1} \pi_{ij} c(i) \ubar{p}_j\right)_{j \in [n]}$.
\end{lemma}

\proof{Proof of Lemma~\ref{wcc}.} \label{proof1} For each $i^* \in \lbrace 0, \ldots, n+1 \rbrace$, we use the idle time reformulation from Lemma~\ref{idletimeref}:
\begin{align*}
& \sum_{i=1}^{i^*} c_i(A_i - \ubar{C}_{i-1}(\pi, A, \lambda))^+ + c_o\left( A_{i^*} + \sum_{k = i^*}^n \pi_{kj}\lambda_j\bar{p}_j - L \right)^+ \\
= & \sum_{l = 1}^{r-1} (c_{i_l} - c_{i_{l+1}}) A_{i_l} + c_{i_r}A_{i_r} - c_{i_1}A_{i_0} - \sum_{l = 1}^r {c_{i_l}} \sum_{{\sigma}=i_{l-1}}^{i_l-1} {\lambda}_{{\sigma}} \ubar{p}_{{\sigma}} + c_o\left( A_{i^*} + \sum_{k = i^*}^n \pi_{kj}\lambda_j\bar{p}_j - L \right)^+
\end{align*} 
Given an arbitrary but fixed sequence and appointment times, it is sufficient to consider only $\left(- \sum_{l = 1}^r {c_{i_l}} \sum_{{\sigma}=i_{l-1}}^{i_l-1} {\lambda}_{{\sigma}} \ubar{p}_{{\sigma}}\right ) + c_o\left( A_{i^*} + \sum_{k = i^*}^n \pi_{kj}\lambda_j\bar{p}_j - L \right)^+$ when maximizing the cost function. Since the first summand is non-positive and the second summand is non-negative, the cost function is maximized by selecting customers such that those with the longest service times show up in the second summand, while, if necessary due to $k > n - i^* + 1$, those with the shortest service times show up in the first summand. Thus, a no-show scenario ${\lambda} \in \Lambda_k$ maximizes the total cost if ${\lambda}_j = 1 $ for some $j \in [n]$ implies the following: If $k \leq n - i^* + 1$, then $ \sum_{i=i^*}^n \pi_{ij} \bar{p}_j$ is among the $k$ largest components of $\left(\sum_{i=i^*}^{n} \pi_{ij} \bar{p}_j\right)_{j \in [n]}$. If $k > n - i^* + 1$, then it holds either $\pi_{ij} = 1$ for some $i \geq i^*$ or $ \sum_{i=1}^{i^*-1} \pi_{ij} c(i) \ubar{p}_j$ is among the $k - (n - i^* + 1)$ smallest components of $\left(\sum_{i=1}^{i^* -1} \pi_{ij} c(i) \ubar{p}_j\right)_{j \in [n]}$.
 \rightline{\qed}
\endproof \vspace{10pt}

To reformulate this result as a mixed-integer linear program, we first introduce auxiliary binary variables $\lambda, \pi(c) , w$. Each of these variables has an index $i^*$ for the worst-case scenario described in Lemma~\ref{wcc}. Now, $\lambda_{i^*ij}$ takes the value 1 if and only if customer $j$ is assigned to the $i$th appointment and shows up. The variable $\pi(c)_{i^*jl}$ takes the value 1 if and only if the $j$th customer is assigned idle time cost $c_l$. The variable $w_{i^*j_1 j_2}$ takes the value 1 if and only if both customer $j_1$ and $j_2$ are assigned to the first $i^* - 1$ appointments, $j_2$ shows up, and $j_1$ is a no-show.

Secondly, for any given but fixed $i^* \in \{0, \ldots, n+ 1\}$, we compute each appointment $i$'s cost function $c(i)$ from Lemma~\ref{wcc} via $c(i) = \sum_{l = 1}^{n+1} \gamma_{il}c_l$ with $c_{n+1} := \max_{l \in [n]}{c_l} $ using any feasible solution $\gamma \in \{0,1\}^{n \times (n + 1)}$ of the following mixed-integer linear problem. 
\begin{align} \label{helo}
\gamma_{il} & \leq \Delta_lM  & \forall l \in [n] \forall i \in[n+1] \\
\sum_{\alpha=i}^{l-1} \Delta_{\alpha} &\leq L(1-\gamma_{il})  & \forall \alpha \in [n] \forall i \in[n+1] \\
\sum_{l=1}^{i-1} \gamma_{il} &= 0 & \forall i \in [n] \\
\sum_{l=i}^{n+1} \gamma_{il} &= 1 & \forall i \in [n]  \label{21}
\end{align}

Constraints~\hyperref[helo]{(54)} ensure that, if there is no positive amount of idle time before an appointment $l \in [n]$, then $\gamma_{il} = 0$. We carefully choose $M \geq \frac{1}{\min_{l\in[n]} \{\Delta_l: \Delta_l> 0 \}}$ to ensure numerical stability. An alternative approach involves replacing (\ref{helo}) with an indicator constraint, namely $\Delta_l = 0 \Rightarrow \gamma_{il} = 0$. Constraints~\hyperref[helo]{(56)}and \hyperref[helo]{(57)} ensure that each appointment $i$ is assigned a unit idle time cost $c_l$ with $l \geq i$. Constraints~\hyperref[helo]{(55)} ensure that, if $\gamma_{i\alpha}=1$, then there occurs no idle time before any of the appointments $i, ..., \alpha -1$. Therefore, Constraints~\hyperref[helo]{(54)} to \hyperref[helo]{(57)} ensure that $\gamma_{il}=1$ holds if and only if $l \in [i,n]$ is the next appointment before which a positive amount of idle time occurs, and $\gamma_{i,n+1}=1$ if there is no such appointment. 

\paragraph{MILP.}
Hence, we have derived the following mixed-integer formulation for the general Robust Appointment Scheduling Problem with Waiting Time Guarantees and $k$ show-ups (RASWTG):
\begin{align} \label{RASWTG-k}
\min\, & U & \\
\text{s.t.}\, & U \geq \sum_{i = 2}^{i^*} c_i \Delta_{i^* i} + c_o \sigma_{i^*} & \forall {i^*} \in [n+1] \\
& \Delta_{i^* i} \geq {A}_i - \ubar{C}_{i^* i-1}   &   \forall {i^*} , i \in [n+1] \\
& \ubar{C}_{i^* i}  = \max\left\lbrace A_i, \ubar{C}_{i^* i-1}  \right\rbrace + \sum_{j=1}^n \lambda_{i^*ij}\ubar{p}_{j} &   \, \forall {i^*}  \in [n+1] \forall i \in [n]  \label{nonlin} \\ 
& \sigma_{i^*} \geq A_{i^*} + \sum_{k = i^*}^n \sum_{j=1}^n \lambda_{i^*kj}\bar{p}_j - L & \forall i^* \in [n] \\
& \lambda_{i^*ij} \leq \pi_{ij} & \forall i^* \in [n+1] \forall i, j \in [n] \\
&  \sum_{i=1}^n \sum_{j=1}^n \lambda_{i^*ij} = k & \forall i^* \in [n+1] \\
& \sum_{j=1}^n \lambda_{i^*ij}  = 1 & \forall {i^*}  \in [n + 1 - k, n+1] \forall i \in [i^*,n]  \\
& \sum_{j=1}^n \lambda_{i^*ij}  = 0 & \forall {i^*}  \in [n-k] \forall i \in [i^*,n] \\
& \sum_{i=1}^n \lambda_{i^*ij_2} - \sum_{i=1}^n \lambda_{i^*ij_1} \leq w_{i^*j_1j_2} &  \forall {i^*}  \in [n+1] \forall j_1,j_2 \in [n] \\
& w_{i^*j_1j_2} \leq 1 - \sum_{i=1}^n \lambda_{i^*ij_1}, \sum_{i=1}^n \lambda_{i^*ij_2} &  \forall {i^*}  \in [n + 1 ] \forall j_1,j_2 \in [n]  \\
& \sum_{l=1}^{n+1} c_l(\pi(c)_{i^*j_2l}\ubar{p}_{j_2} -\pi(c)_{i^*j_1l}\ubar{p}_{j_1} ) \leq M(1-w_{i^*j_1j_2}) & \forall {i^*}  \in [n + 1 - k, n+1] \forall j_1,j_2 \in [n]   \\
& \bar{p}_{j_2} - \bar{p}_{j_1} \leq M(1-w_{i^*j_1j_2}) & \forall {i^*}  \in [n - k ] \forall j_1,j_2 \in [n] \\
& (\ref{helo}) - (\ref{21}) \text{ with additional index } i^* \in [n+1] \\
& - 1 + \pi_{ij} + \gamma_{i^*il} \leq \pi(c)_{i^*jl} & \forall i^* \in [n+1] \forall j, l \in[n] \\
& A_{l} + \sum\limits_{k = l}^{i-1} \sum_{j = 1}^n \pi_{kj} \bar{p}_{j} - A_{i} \leq \sum_{j=1}^n \pi_{ij}W_j & \forall i, l: i \in [n] , l \in[i-1] , i - l \leq k-1 \\
& A_{l} + (k-1)\,\alpha_{i,l} + \sum_{\sigma = l}^{i-1}p_{il\sigma} - A_{i} \leq \sum_{j=1}^n \pi_{ij}W_j & \forall i \in [n] , l \in[i-1] : i - l > k-1 \\
& \alpha_{il} + p_{il\sigma} \geq \sum_{j=1}^n \pi_{\sigma j}\bar{p}_j & \forall i,l,\sigma \in [n] \\
& \sum_{j=1}^n \pi_{ij} = 1, \sum_{i=1}^n \pi_{ij} = 1, \sum_{j=1}^n \pi(c)_{i^*jl} = 1 & \forall i,j,l \in [n] \\
&A_1 = 0, \ubar{C}_{i^*0} = 0, A_{n+1} = L & \forall {i^*}  \in [n+1]  \\
& \Delta_{i^*i} , \Delta_{n+1,i}, \sigma_{i^*}, \alpha_{i,l} , p_{il\sigma} \geq 0 & \forall {i^*}  \in [n+1] \forall i,l,\sigma \in [n] \\
& \lambda_{i^*ij},  \pi(c)_{i^*jl}, \gamma_{il}^{(i^*)}, w_{i^*j_1j_2}  \in \lbrace 0,1 \rbrace & \forall {i^*}  \in [n+1]  \forall i,j,j_1, j_2, l \in [n] \\
& A_i \geq 0, \pi_{ij} \in \lbrace0,1\rbrace & \forall i,j \in [n]
\end{align}

The Objective Function~\hyperref[RASWTG-k]{(58)} minimizes the upper bound on the total cost, which is computed in Constraints~\hyperref[RASWTG-k]{(59) -- (71)}. Specifically, Constraint~\hyperref[RASWTG-k]{(59)} sets the upper bound for cost incurred from idle time $\Delta_{i^*i}$ and overtime $\sigma_{i^*}$ in each worst-case scenario corresponding to $i^* \in [n + 1]$ (see Lemma~\ref{wcc}). Constraint~\hyperref[RASWTG-k]{(60) -- (61)} define the worst-case idle time before the $i^*$th appointment as a function of $\lambda$. Similarly, Constraint~\hyperref[RASWTG-k]{(62)} defines the worst-case overtime accumulated starting from the $i^*$th appointment. Constraint~\hyperref[RASWTG-k]{(64)} ensures that a total of $k$ customers show up. 
The following constraints implement the results from Theorem~\ref{wc}. Constraints~\hyperref[RASWTG-k]{(65)} ensure that if $k \geq n - i^* + 1$, then all customers schedules for the $k - (n - i^* + 1)$ last appointments show up. Constraint~\hyperref[RASWTG-k]{(66)} allows customers to show up for the first $i^* - 1$ appointments only if $k > n - i^* +1$. Constraints~\hyperref[RASWTG-k]{(67) -- (70)} determine that among the first $i^*$ appointments, only those with the smallest product $c_l\ubar{p}_j$ show up, and among the subsequent appointments those with the longest maximum service times. Finally, Constraints~\hyperref[RASWTG-k]{(72) -- (74)} ensure the waiting time guarantees as derived in (\ref{waitnoshows}). 
\vspace{10pt}

\section{Optimal Interappointment Times}
 \label{intapptimes} 

Optimal schedules are computed using Gurobi for a case with ten identical customers, each with a service time ranging from 15 to 25 minutes and a waiting time guarantee of 30 minutes. The resulting optimal interappointment times are shown in Figure~\ref{dome1}.

\begin{figure}[h]
\caption{Optimal Interappointment Times \normalfont{(in Minutes)}}\label{dome1}
\begin{subfigure}{5cm}
\caption{\EGT\rm Decreasing Idle Time Costs}
\begin{tikzpicture}[scale = 1, font=\EGT\rm]
\begin{axis}[ ylabel={\EGT\rm },
xmax=10.5, ymin = 0, ymax=35,xtick={1,2,3,4,5,6,7,8,9}, xticklabels = {$A_2-A_1$, $A_3-A_2$, $A_4-A_3$,$A_5-A_4$,$A_6-A_5$,$A_7-A_6$,$A_8-A_7$,$A_9-A_8$,$A_10-A_9$}, xticklabel style={rotate=90} , ytick={0,10,20,30},legend pos=north west,mark size=0.25pt,mark=o,no markers, axis lines = left,ylabel style={align=center},
every axis plot/.append style={thick},width=5cm,height=3cm]
\addplot[TUMalblack,dotted,every mark/.append style={fill=blue!80!black}] table[x index=0,y index=1,col sep=comma] {data/data30.dat};
\addplot[TUMalblack,every mark/.append style={fill=blue!80!black}] table[x index=0,y index=2,col sep=comma] {data/data30.dat};
\end{axis}
\end{tikzpicture}
\end{subfigure}
\begin{subfigure}{5cm}
\caption{\EGT\rm Constant Idle Time Costs}
\begin{tikzpicture}[scale = 1, font=\EGT\rm]
\begin{axis}[ ylabel={\EGT\rm },
xmax=10.5, ymin = 0, ymax=35,xtick={1,2,3,4,5,6,7,8,9}, xticklabels = {$A_2-A_1$, $A_3-A_2$, $A_4-A_3$,$A_5-A_4$,$A_6-A_5$,$A_7-A_6$,$A_8-A_7$,$A_9-A_8$,$A_10-A_9$} , xticklabel style={rotate=90} , ytick={0,10,20,30},legend pos=north west,mark size=0.25pt,mark=o,no markers, axis lines = left,ylabel style={align=center},
every axis plot/.append style={thick},width=5cm,height=3cm]
\addplot[TUMalblack,dotted,every mark/.append style={fill=blue!80!black}] table[x index=0,y index=3,col sep=comma] {data/data30.dat};
\addplot[TUMalblack, every mark/.append style={fill=blue!80!black}] table[x index=0,y index=4,col sep=comma] {data/data30.dat};
\end{axis}
\end{tikzpicture}
\end{subfigure}
\begin{subfigure}{5cm}
\caption{\EGT\rm Increasing Idle Time Costs}
\begin{tikzpicture}[scale = 1, font=\EGT\rm]
\begin{axis}[ylabel={\EGT\rm },
xmax=10.5, ymin = 0, ymax=55,xtick={1,2,3,4,5,6,7,8,9}, xticklabels = {$A_2-A_1$, $A_3-A_2$, $A_4-A_3$,$A_5-A_4$,$A_6-A_5$,$A_7-A_6$,$A_8-A_7$,$A_9-A_8$,$A_10-A_9$}, xticklabel style={rotate=90} , ytick={0,10,20,30, 40, 50},legend pos=north west,mark size=0.25pt,mark=o,no markers, axis lines = left,ylabel style={align=center},
every axis plot/.append style={thick},width=5cm,height=3cm]
\addplot[TUMalblack,dotted,every mark/.append style={fill=blue!80!black}] table[x index=0,y index=5,col sep=comma] {data/data30.dat};
\addplot[TUMalblack, every mark/.append style={fill=blue!80!black}] table[x index=0,y index=6,col sep=comma] {data/data30.dat};
\end{axis}
\end{tikzpicture}
\end{subfigure}

\begin{subfigure}{1cm}
\caption*{\EGT\rm }
{\begin{tikzpicture}[scale = 1, font=\EGT\rm]
\end{tikzpicture}}
\end{subfigure}
\begin{subfigure}{0.5cm}
\begin{tikzpicture}[remember picture, overlay, transform shape]
\node
at ($(13.6,-0.5)$)
{
{\begin{tikzpicture}[scale = 1, font=\EGT\rm]
\begin{customlegend}[legend entries={\EGT\rm 0\% No-Shows,20\% No-Shows}]
\addlegendimage{TUMalblack,draw=black, thick}
\addlegendimage{TUMalblack,draw=black, thick,dotted}
\end{customlegend}
\end{tikzpicture}}
};
\end{tikzpicture}
\end{subfigure}
\end{figure}
 
In the case of zero no-shows, the ASAP scheduling rule in Theorem~\ref{opt-times} implies increasing or varying interappointment times at the beginning of the scheduling horizon to avoid early idle times, followed by a constant pattern (limited by the waiting time guarantees). In this sense, the semi-plateau-dome pattern is a customer-driven variant of the plateau-dome-shaped pattern for zero-no shows.

Moving on to the case that only $k < n$ patients show up, no-shows do not affect the worst-case completion times of the first~$k$~appointments, but they affect the worst-case completion times of later appointments. In particular, some later interappointment times can be shortened in favor of lower cost. Due to the adherence to the waiting time guarantees, the interappointment times rise again to their previous level afterward. This observation leads to spikes typically after $k$ customers. We consider this structure a customer-driven variant of the plateau-dome-shaped pattern with no-shows. 

\vspace{10pt}

\section{Computational Study Details} \label{A_comp}

\subsection{Instances}

\begin{figure}[b]
\caption{Service Times by Examination Type \normalfont{(Number of Observations in Parentheses)}} \label{estimates}
\begin{tikzpicture}[font=\EGT\rm]
\begin{axis}
[height=6.0cm, width=12.0cm,draw=TUMalblack, xlabel = Service time (in minutes), 
ytick={1,2,3,4,5,6,7,8,9,10},
yticklabels={Thorax (2.197), Throat (832), 
Abdomen (476), Whole body (168), Pancreas (155), Brain skull (123), Knee (114)},
]
\addplot+[TUMalblack,dash pattern = on 3pt off 0pt, mark=*, mark size = 1.0pt, mark options = {fill = white},
boxplot prepared={
upper quartile=10.06,
lower quartile=7.14,
upper whisker=24.31,
lower whisker=3.09,
median = 8.13
},
] coordinates {
 (0,25.46) (0,28.01) (0,28.45) (0,30.41) (0,30.50) (0,31.26) (0,35.28) (0,37.53) (0, 38.33) (0,43.15) (0,53.14)};
\addplot+[TUMalblack,dash pattern = on 3pt off 0pt,mark=*,
   mark size = 1.0pt, mark options = {fill = white},
boxplot prepared={
upper quartile=11.14,
lower quartile=8.18,
upper whisker=25.09, 
lower whisker=4.10,
median = 9.32,
},
] coordinates {
 (0,29.50) (0,31.33) (0,33.20) (0,42.50)};
\addplot+[TUMalblack,dash pattern = on 3pt off 0pt,mark=square*,
   mark size = 1.0pt, mark options = {fill = white},
boxplot prepared={
upper quartile=10.38,
lower quartile=4.59,
upper whisker=32.46, 
lower whisker=2.34,
median = 7.07,
},
] coordinates {
(0, 37.03) (0, 38.43) 
(0, 45.26) (0, 45.38) (0, 60.13) (0, 90.55)};
\addplot+[TUMalblack,dash pattern = on 3pt off 0pt,mark=square*,
   mark size = 1.0pt, mark options = {fill = white},
boxplot prepared={
upper quartile=21.57,
lower quartile=14.00,
upper whisker=32.46, 
lower whisker=13.29,
median = 17.26,
},
] coordinates {
(0, 36.08) (0, 40.27)};
\addplot+[TUMalblack,dash pattern = on 3pt off 0pt,mark=square*,
   mark size = 1.0pt, mark options = {fill = white},
boxplot prepared={
upper quartile=10.55,
lower quartile=7.15,
upper whisker=15.40, 
lower whisker=5.22,
median = 8.27,
},
] coordinates {
(0, 17.57) (0, 18.54) (0, 19.57) (0, 20.01) (0, 20.02) (0, 20.55)};
\addplot+[TUMalblack,dash pattern = on 3pt off 0pt,mark=square*,
   mark size = 1.0pt, mark options = {fill = white},
boxplot prepared={
upper quartile=11.21,
lower quartile=6.08,
upper whisker=17.53, 
lower whisker=3.30,
median = 7.30,
},
] coordinates {
(0, 19.37) (0, 20.21) (0, 22.02) (0, 22.50) };
\addplot+[TUMalblack,dash pattern = on 3pt off 0pt,mark=square*,
   mark size = 1.0pt, mark options = {fill = white},
boxplot prepared={
upper quartile=10.03,
lower quartile=6.41,
upper whisker=17.55, 
lower whisker=4.55,
median = 8.06,
},
] coordinates {
(0, 21.17) (0, 22.27) (0, 23.52) (0, 23.59) (0,24.04) (0, 25.25) (0, 26.59) (0, 28.23) (0, 176.43) };
\end{axis}
\end{tikzpicture}
\end{figure}
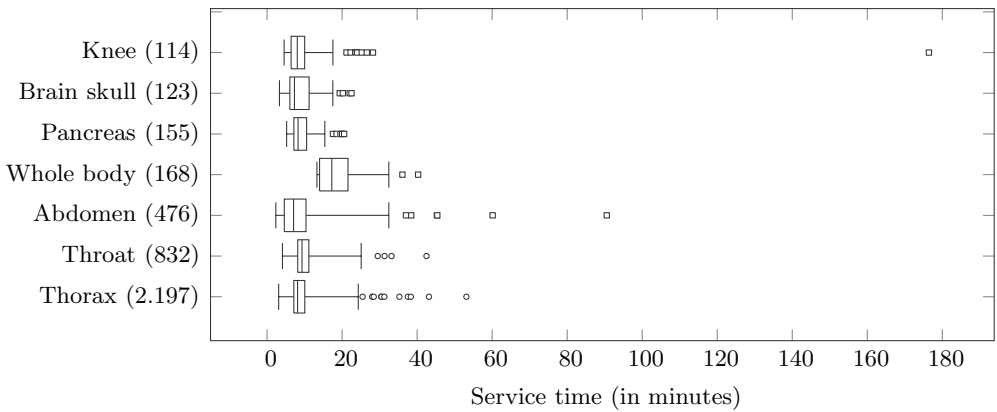
%


\label{instanceGen}
The chosen patient numbers, $n \in \lbrace 5, 10, 20 \rbrace$, are aligned with our daily CT examination records and are consistent with previous computational studies utilizing distributionally robust optimization approaches \citep{qi2017mitigating, jiang2017integer, kong2020appointment}. To capture the variability in real-world scenarios, and following consultation with radiology management, we introduce variations in no-show rates across 0\%, 10\%, and 20\%, achieved by setting $k \in \lbrace n,\lfloor 0.9n \rfloor,\lfloor 0.8n \rfloor \rbrace$. Figure~\ref{estimates} presents a boxplot illustrating the service times for the most common examination types. 


\subsection{Sensitivity Analysis: The Number of Customers}
 \label{remark_no}
As can be seen in Table \ref{tab:horizon}, the RASWTG approach leads to shorter waiting times and longer idle times as the number of scheduled customers increases. Specifically, the ASAP scheduling rule from Theorem~\ref{opt-times}, which is optimal for constant and decreasing costs, provides a rationale for this trend:

The first few appointments can all be scheduled at the start time of the scheduling horizon, as long as the sum of their maximum service times does not exceed the waiting time guarantee of the subsequent appointment. For these appointments, waiting times accumulate in proportion to the service times realized. For all subsequent appointments, the start time can be calculated as the worst-case sum of the previous service times minus the waiting time guarantee. As the number of customers treated increases, we consider it increasingly unlikely that all customers require their maximum service time. Consequently, we expect short waiting times and longer idle times before late appointments. 


We note that this trend could be mitigated by using uncertainty budgets that control the number of customers who require maximum service time. 

\subsection{Comparison of RASWTG and Sample Average Approximation}\label{SAA_app}


\textcolor{black}{\hyperref[RASWTG-k]{RASWTG} minimizes worst-case costs from idle time and overtime while ensuring worst-case waiting time guarantees. This worst-case optimization approach is selected to adequately account for individual patient experiences (see Section~\ref{subSection-literature}) and to mitigate excessive idle time and overtime. However, it may lead to higher total realized costs compared to minimizing expected costs. To evaluate this effect, we compare waiting times, idle time, overtime, and total cost of \hyperref[RASWTG-k]{RASWTG} with those from a sample average approximation (SAA) approach. Unless stated otherwise, we use \hyperref[RASWTG-k]{RASWTG} in the base case setting. For each patient~$j$, we draw $N \in \{10, 25\}$ service time samples $p_j^1, \ldots, p_j^N$ based on the examination type. The resulting SAA with Robust Waiting Time Guarantees (SAA-RWTG) is formulated as the following MILP.}


\begin{align} \label{SAA-RWTG} \color{TUMblue}
\min \, & \frac{1}{N} \sum_{l=1} ^N U_l & \\
\text{s.t.}\,
& U_l \geq \sum_{i = 1}^{n+1} c_i \Delta_{l,i} + c_o \sigma_{l} & \forall {l} \in [N] \label{objSAA} \\
& \Delta_{l,i} \geq {A}_i -  {C}_{l,i-1} & \forall i \in [n+1] \forall {l} \in [N] \label{idleSAA} \\
& \sigma_{l} \geq  {C}_{l, n}  - L &  \forall {l} \in [N] \label{overSAA} \\
&  {C}_{l,i} = \max\left\lbrace A_i,  {C}_{l, i-1} \right\rbrace + \sum_{j=1}^n \pi_{ij} {p}_{j}^l & \, \forall i \in [n]  \forall {l} \in [N] \label{nonlinSAA} \\
& A_{l} + \sum\limits_{k = l}^{i-1} \sum_{j = 1}^n \pi_{kj} \bar{p}_{j} - A_{i} \leq \sum_{j=1}^n \pi_{ij}W_j & \forall i, l: i \in [n] , l \in[i-1] \label{WTGSAA} \\
& \sum_{j=1}^n \pi_{ij} = 1, \sum_{i=1}^n \pi_{ij} = 1 & \forall i,j \in [n] \label{seqSAA} \\
& A_1 = 0,  {C}_{l,0} = 0, A_{n+1} = L  & \forall {l} \in [N] \label{2-0nsSAA} \\
& A_i,  {C}_{l,i} \geq 0 & \forall {i} \in [n] \\
& \Delta_{l,i}, \sigma_{l} \geq 0 & \forall {i} \in [n+1]  \forall {l} \in [N] \\
& \pi_{ij} \in \lbrace 0,1 \rbrace & \forall i,j \in [n]
\end{align}

 \textcolor{black}{
The Objective Function~(\ref{SAA-RWTG}) minimizes the upper bound on total cost, as specified in Constraint~(\ref{objSAA}). Constraint~(\ref{idleSAA}) upper bounds sample-dependent idle times, while Constraint~(\ref{overSAA}) upper bounds for overtime. Completion times are defined in Constraint~(\ref{nonlinSAA}), and the waiting-time and assignment constraints are captured by Constraints~(\ref{WTGSAA}) and~(\ref{seqSAA}), respectively, as in the Robust Appointment Scheduling Problem with Waiting Time Guarantees. Table~\ref{tab:costSAA} summarizes the results explained in Section~\ref{choice_uncertaintyset}.}

\begin{table}[h]  
\caption{Comparison of SAA-RWTG and RASWTG}
\centering \footnotesize
\begin{tabular}{lrrrrrrrrrrrrr}
\hline
&& \multicolumn{3}{c}{\hyperref[SAA-RWTG]{SAA-RWTG}} &&  {RASWTG} 
\\
&& $N = 10$ &&  $N = 25$ && 
\\ \cmidrule{3-3} \cmidrule{5-5} \cmidrule{7-7} 
{Share of patients waiting at most 30 minutes:} 
&& 96.01\%  && 95.67\% &&  {96.88\%}
\\
{Idle time up to the last appointment:} 
&& 17:38  && 16:56 &&  {17:11}
\\
{Total idle time:} 
&& 37:41 && 37:30  &&  {37:40} \\
{Overtime:} 
&& 1:47  && 1:36 &&  {1:46}  \\
Total cost: 
&& 39.92 &&  39.49 &&   {39.86} \\ 
Average runtime (seconds): && 111.21 && 1060.44 && 0.00 \\ \hline
\end{tabular}
\label{tab:costSAA}
\end{table}

\textcolor{black}{
}



\subsection{Utilitarian Approach}

\begin{figure}[b]
\caption{Optimal Interappointment Times  \normalfont{(in Minutes)}}\label{dome}
\begin{subfigure}{5cm}
\caption{\normalfont{WSRAS with $c_w = 0.00001$}}
\begin{tikzpicture}[scale = 1, font=\EGT\rm]
\begin{axis}[ ylabel={\EGT\rm },
xmax=10.5, ymin = 0, ymax=70,xtick={1,2,3,4,5,6,7,8,9}, xticklabels = {$A_2-A_1$, $A_3-A_2$, $A_4-A_3$,$A_5-A_4$,$A_6-A_5$,$A_7-A_6$,$A_8-A_7$,$A_9-A_8$,$A_10-A_9$}, xticklabel style={rotate=90} , ytick={0,10,20,30, 40, 50, 60, 70},legend pos=north west,mark size=0.25pt,mark=o,no markers, axis lines = left,ylabel style={align=center},
every axis plot/.append style={thick},width=5cm,height=4cm]
\addplot[TUMalblack,dashed,every mark/.append style={fill=blue!80!black}] table[x index=0,y index=1,col sep=comma] {data/data31.dat};
\addplot[TUMalblack,every mark/.append style={fill=blue!80!black}] table[x index=0,y index=4,col sep=comma] {data/data30.dat};
\end{axis}
\end{tikzpicture}
\end{subfigure}
\begin{subfigure}{5cm}
\caption{\normalfont{WSRAS with $c_w = 0.1$}}
\begin{tikzpicture}[scale = 1, font=\EGT\rm]
\begin{axis}[ ylabel={\EGT\rm },
xmax=10.5, ymin = 0, ymax=70,xtick={1,2,3,4,5,6,7,8,9}, xticklabels = {$A_2-A_1$, $A_3-A_2$, $A_4-A_3$,$A_5-A_4$,$A_6-A_5$,$A_7-A_6$,$A_8-A_7$,$A_9-A_8$,$A_10-A_9$}, xticklabel style={rotate=90} , ytick={0,10,20,30, 40, 50, 60, 70},legend pos=north west,mark size=0.25pt,mark=o,no markers, axis lines = left,ylabel style={align=center},
every axis plot/.append style={thick},width=5cm,height=4cm]
\addplot[TUMalblack,dashed, every mark/.append style={fill=blue!80!black}] table[x index=0,y index=2,col sep=comma] {data/data31.dat};
\addplot[TUMalblack,every mark/.append style={fill=blue!80!black}] table[x index=0,y index=4,col sep=comma] {data/data30.dat};
\end{axis}
\end{tikzpicture}
\end{subfigure}
\begin{subfigure}{5cm}
\caption{\normalfont{WSRAS with $c_w = 1$}}
\begin{tikzpicture}[scale = 1, font=\EGT\rm]
\begin{axis}[ylabel={\EGT\rm },
xmax=10.5, ymin = 0, ymax=70,xtick={1,2,3,4,5,6,7,8,9}, xticklabels = {$A_2-A_1$, $A_3-A_2$, $A_4-A_3$,$A_5-A_4$,$A_6-A_5$,$A_7-A_6$,$A_8-A_7$,$A_9-A_8$,$A_10-A_9$}, xticklabel style={rotate=90} , ytick={0,10,20,30, 40, 50, 60, 70},legend pos=north west,mark size=0.25pt,mark=o,no markers, axis lines = left,ylabel style={align=center},
every axis plot/.append style={thick},width=5cm,height=4cm]
\addplot[TUMalblack,dashed, every mark/.append style={fill=blue!80!black}] table[x index=0,y index=3,col sep=comma] {data/data31.dat};
\addplot[TUMalblack,every mark/.append style={fill=blue!80!black}] table[x index=0,y index=4,col sep=comma] {data/data30.dat};
\end{axis}
\end{tikzpicture}
\end{subfigure}

\begin{subfigure}{15cm}
\begin{tikzpicture}[remember picture, overlay, transform shape]
\node
at ($(14,-0.75)$)
{
{\begin{tikzpicture}[scale = 1, font=\EGT\rm]
\begin{customlegend}[legend entries={\EGT\rm WSRAS, \hyperref[RASWTG-k]{RASWTG}}]
\addlegendimage{TUMalblack,draw=black, thick,dashed}
\addlegendimage{TUMalblack,draw=black, thick}
\end{customlegend}
\end{tikzpicture}}
};
\end{tikzpicture}
\end{subfigure}
\end{figure}
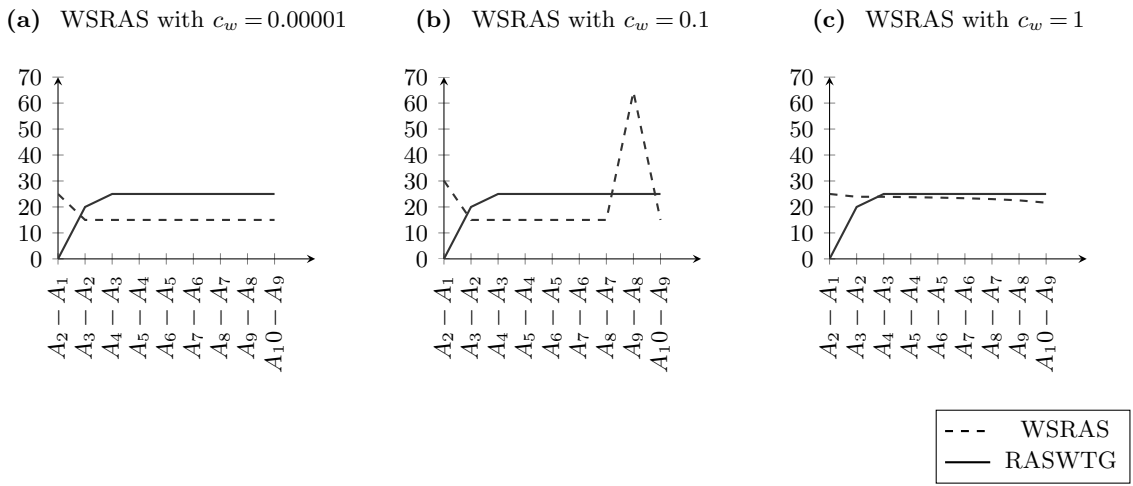

We briefly explore optimal interappointment times to enhance our understanding of optimal WSRAS schedules, as illustrated in Figure~\ref{dome}. We again compute the interappointment times for ten identical customers with a minimum service time of 15 minutes, a maximum service time of 25 minutes, and a waiting time guarantee of 30 minutes. In contrast to \hyperref[RASWTG-k]{RASWTG}, WSRAS interappointment times start at a higher level. Then, depending on the waiting time weight, they gradually decrease until a certain point. For small waiting time weights, most interappointment times match the minimum service time of 15 minutes, leading to little idle time but exploding waiting times. Conversely, higher waiting time weights yield interappointment times closer to the maximum service time of 25 minutes, resulting in short waiting times but extended idle time. Intriguingly, when $c_w = 0.1$, WSRAS accumulates waiting times until, at a certain point, they are alleviated by a single long interappointment time.

Finally, we note that the \hyperref[TRAS]{WSRAS} problem with constant idle time costs closely resembles problem (1) in \cite{schulz2019robust}, except that the latter does not consider a fixed planning horizon $L$. While initially minimizing a weighted sum of idle times and waiting time costs under a collective worst-case scenario, \cite{schulz2019robust} demonstrate the sufficiency of considering the same $n + 1$ scenarios as in \hyperref[TRAS]{WSRAS}.

Moreover, comparing an \hyperref[RASWTG]{RASWTG}-optimal schedule that does not consider a fixed planning horizon (e.g., $c_{n+1} = c_o = 0$) to a schedule computed with the optimal scheduling rule in \cite{schulz2019robust} and under the respective assumptions, we find that \hyperref[RASWTG]{RASWTG} improves capacity utilization for sufficiently long waiting time guarantees. In particular, it is easy to show that the \hyperref[RASWTG]{RASWTG} schedule reduces worst-case idle time costs in the case of $ W > \sum_{i=1}^{n-1}\sum_{j=1}^n \pi_{ij} \frac{u}{u + o_{\geq i}} \left( \bar{p}_{j} - \ubar{p}_{j} \right)$.

 \end{APPENDIX}


\end{document}